\documentclass[final,times,onecolumn]{elsarticle}
\usepackage[a4paper]{geometry}
\usepackage{setspace} 
\usepackage{graphicx}
\usepackage{amssymb}
\usepackage{newtxtext}
\usepackage{newtxmath}
\usepackage{natbib}
\usepackage{xcolor}
\usepackage{hyperref}
\usepackage{cleveref} 
\usepackage{float}
\usepackage{threeparttable}

\usepackage{amsmath}
\usepackage{booktabs}
\usepackage{tabularx}
\usepackage{array} 
\usepackage{caption}  
\usepackage{algorithm} 
\usepackage{algpseudocode}
\hypersetup{
    colorlinks=true,
    citecolor=blue,    
    linkcolor=blue,    
    filecolor=blue,    
    urlcolor=blue      
}

\usepackage[table]{xcolor} 
\usepackage{array} 

\crefname{equation}{equation}{Eqs.}
\Crefname{equation}{equation}{Eqs.}
\crefname{figure}{figure}{figures}
\Crefname{Figure}{Figure}{Figures}
\crefname{algorithm}{Algorithm}{Algorithms}
\Crefname{algorithm}{Algorithm}{Algorithms}
\crefname{section}{section}{sections}
\Crefname{Section}{Section}{Sections}
\crefname{table}{table}{tables}
\Crefname{Table}{Table}{Tables} 

\usepackage[normalem]{ulem}

\begin{document}

\begin{frontmatter}

\title{Optimizing synthetic jet placement for control stability in increasing reynolds number flows based on deep reinforcement learning}

\author{Wang Jia, Hang Xu}

\affiliation{organization={State Key Laboratory of Ocean Engineering, School of Ocean and Civil Engineering, Shanghai Jiao Tong University},
            city={Shanghai},
            postcode={200240}, 
            country={China}}
        
\begin{abstract}
This study utilizes reinforcement learning algorithms to develop advanced control strategies for manipulating synthetic jets positioned on a cylinder, aiming to achieve effective flow control. It addresses critical challenges in achieving robust performance across moderate to high Reynolds numbers ($Re$) and varying jet placement configurations. 
The results demonstrate that placing synthetic jets near flow separation regions yields optimal control performance. At $Re = 500$, the multi-objective flow control strategy achieves up to 37.4\% drag reduction and 99\% suppression of lift oscillations, while completely suppressing vortex shedding. At $Re = 1,000$, drag reduction improves to 49.1\%, with equally effective lift suppression and complete vortex shedding suppression. At $Re = 2,000$, the strategy achieves a maximum drag reduction of 58.8\% and suppresses 90\% of lift oscillations, although some residual wake instabilities persist.
Additionally, synthetic jets effectively disrupted detached vortices and stabilized the wake, resulting in streamlined flow and suppressed vortex shedding.
These findings underscore the pivotal influence of jet placement in optimizing flow control efficacy, while affirming the advanced capability of control strategies to tackle intricate  challenges.
\end{abstract}


\end{frontmatter}

\section{INTRODUCTION}

The effectiveness of deep reinforcement learning (DRL) in active flow control (AFC) has been extensively demonstrated, with prior studies focusing on the robustness of control strategies \cite{rabault2019artificial,dong2023surrogate,jia2024robust,zong2025closed}, the enhancement of control performance \cite{parisRobust,Wang2024,Chatzimanolakis}, and the optimization of control algorithms \cite{wang2024learn,jia2024optimal,jia2024jetsactuator}. 
An intriguing and highly challenging research direction involves investigating the impact of synthetic jet placement on control performance and exploring the feasibility of achieving optimal control performance at medium and high $Re$. 
To address these objectives, two critical challenges must be tackled:
(i) understanding the sensitivity of control stability to variations in synthetic jet placement, and  
(ii) achieving complete suppression of vortex shedding under the complex dynamics of medium-to-high Reynolds number flows. This study will address and tackle these two challenges.

\subsection{Active flow control}

Since Prandtl introduced boundary layer theory and suction techniques to delay flow separation around cylinders, flow control has become a fundamental research area in fluid mechanics, driving innovation across various domains \cite{schlichting2016boundary}. Flow control strategies are classified into passive approaches, which rely on fixed geometrical modifications without external energy, and AFC, which utilizes external energy to dynamically manipulate flow properties \cite{Ashill2005}. 
AFC enables real-time, high-precision regulation of fluid flow, offering significant advantages such as superior responsiveness and energy efficiency, which have led to extensive applications in aerospace, energy, and transportation sectors to enhance system stability, reduce drag, and suppress adverse flow phenomena \cite{li2025formation}. 
AFC involves two main approaches—open-loop control and closed-loop control. Open-loop control relies on preset inputs without feedback, making it simple but less adaptable to real-time changes, thereby struggling with model errors or external disturbances \cite{LI2024109110}. In contrast, closed-loop control gathers real-time flow field measurement data to dynamically adjust control parameters, exhibiting higher accuracy and robustness under complex flow conditions \cite{LI2022106034}. 

Closed-loop control strategies present significant challenges due to the need for precise sensors, fast-response actuators, and sophisticated algorithms for real-time flow estimation and control \cite{STANEWSKY2001583}. These requirements become particularly demanding in complex flow scenarios, where high computational loads can limit responsiveness and accuracy. Furthermore, the presence of unreliable or noisy sensor data can destabilize the system, compromising overall performance \cite{Bharghava2024}.
In addition to these technical hurdles, the development of effective AFC strategies is further hindered by the complexity of fluid systems governed by nonlinear Navier-Stokes equations and the expansive, high-dimensional control parameter space. External disturbances in physical environments, coupled with imperfections in actuator manufacturing and installation, intensify these challenges. 
Consequently, overcoming these obstacles has made the development of AFC strategies a pivotal area of research in fluid mechanics and control engineering \cite{vinuesa2022flow}.

Recent advances in machine learning have created new opportunities across scientific disciplines \cite{LUO2024118493,feng2024multistation,jiang2025koopman}. In particular, DRL has shown great promise for AFC \cite{rabault2020deep,vinuesa2022flow,jiang2024balanced,zhao2025comparative}.
Unlike traditional reinforcement learning, which relies on manually designed features or simple models that struggle in high-dimensional environments, DRL utilizes deep neural networks to automatically extract features from complex states, effectively addressing the curse of dimensionality \cite{Mnih2015,granter2017alphago}. DRL excels in closed-loop flow control by enabling real-time decision-making and flow state manipulation \cite{zong2024experimental,xiang2024experimental}. Leveraging deep neural networks, it effectively models and optimizes nonlinear flow phenomena, addressing complex control tasks in dynamic environments \cite{zong2020momentum,LI2025,feng2025typhoon,ghaemi2025integ}.

\subsection{Recent advances in the Field}

In 2019, \citeauthor{rabault2019artificial} \citep{rabault2019artificial} integrated reinforcement learning with synthetic jet actuation for cylinder flow control, a pivotal milestone in AFC. 
The study enhances control robustness \citep{dong2024interactive,jia2024robust,ghaemi2025model}, optimizes sensor placement \citep{parisRobust,jiang2023reinforcement}, and integrates physical principles into learning algorithms \citep{liReinforcementlearning}.
Further advances included flow feature extraction \citep{wang2024dynamic} and extending control from two-dimensional to three-dimensional flows \citep{Suarez2024,Chatzimanolakis}. Current flow control strategies remain suboptimal, especially at moderate to high $Re$, where achieving robust, steady control is challenging. For instance, \citeauthor{tangRobust2020} \citep{tangRobust2020} achieved drag reductions of 5.7\% to 38.7\% at $ Re = 100 $ to $ Re = 400 $, but failed to fully suppress vortex shedding, indicated by persistent drag coefficient oscillations. Similarly, \citeauthor{he2023policy} \citep{he2023policy} reported drag reductions of 6.3\% to 23.7\% at $ Re = 100 $ to $ Re = 300 $, yet observed significant fluctuations in drag coefficient and jet velocity. 

Challenges become more pronounced at higher $Re$.
\citeauthor{renApplying} \citep{renApplying} obtained a 30\% drag reduction at $ Re = 1,000 $, but persistent oscillations revealed unresolved flow instabilities. 
\citeauthor{jia2024robust} \cite{jia2024robust} developed robust and adaptive deep reinforcement learning strategies, achieving significant drag reduction ranging from 14.4\% to 47.0\% at $Re = 100$ to 400, respectively. Notably, this is the first work to fully suppress vortex shedding around a square cylinder.
\citeauthor{Wang2024} \cite{Wang2024} utilized deep reinforcement learning to achieve superior flow control performance, particularly for elliptical cylinders with extreme aspect ratios.
Despite progress in DRL-based AFC, which struggles with vortex shedding and flow stabilization, table \ref{tab:table1} highlights the versatility and adaptability of these strategies across various $Re$ in addressing complex challenges.

\begin{table*}[t]
\centering
\caption{Comparison of DRL-based flow control performance for flow around a circular cylinder of diameter D.}
\begin{tabularx}{\textwidth}{
  >{\centering\arraybackslash}p{0.035\textwidth}
  >{\centering\arraybackslash}p{0.06\textwidth}
  >{\centering\arraybackslash}p{0.065\textwidth}
  >{\centering\arraybackslash}p{0.195\textwidth}
  >{\centering\arraybackslash}p{0.082\textwidth}
  >{\centering\arraybackslash}p{0.24\textwidth}
  >{\centering\arraybackslash}p{0.17\textwidth}
}
\toprule
\toprule
    $Re$ & Domain width & Blocking effect & Method & Drag reduction & Control action & Reference \\[3pt]
\midrule
    100   & 4.1D & 24.4 \% & Synthetic Jets &  8\% & 2 jets (1 top \& 1 bottom) & \citet{rabault2019artificial} \\  
    200   & 10 D & 10.0 \%  & Oscillating Cylinder &  8\% & Oscillating amplitude & \citet{jiang2023reinforcement} \\ 
    300   & 20 D & 5.0 \% & Rotational Oscillation &  21\% &  Rotational angular velocity & \citet{HAN2022109008} \\  
    400   & 4.1D & 24.4 \% & Synthetic Jets &  39\% & 4 jets (2 top \& 2 bottom) & \citet{tangRobust2020} \\   
    500   & 4.1D & 24.4 \% & Synthetic Jets &  32\% & 2 jets (1 top \& 1 bottom) & \citet{wang2024dynamic} \\   
    1,000 & 4.1D & 24.4 \% & Synthetic Jets &  20\% & 2 jets (1 top \& 1 bottom) & \citet{Varela2022} \\  
    1,000 & 4.1D & 24.4 \% & Synthetic Jets &  30\% & 2 jets (1 top \& 1 bottom) & \citet{renApplying} \\  
    1,000 & 4.1D & 24.4 \% & Synthetic Jets &  46\% & 2 jets (1 top \& 1 bottom) & \citet{wang2024dynamic} \\       
    2,000 & 4.1D & 24.4 \% & Synthetic Jets &  18\% & 2 jets (1 top \& 1 bottom) & \citet{Varela2022} \\
\bottomrule
\bottomrule
\end{tabularx}
\label{tab:table1}
\end{table*}

The placement of synthetic jets is a critical factor in influencing the effectiveness of AFC, as it directly determines the interaction between the jets and the flow field. Properly positioned jets can enhance control efficiency by targeting regions of high flow instability, such as wake vortices, thereby maximizing drag reduction and lift stabilization. Conversely, suboptimal placement may lead to inefficient energy usage and limited control effects. 
Traditional experimental methods have conducted extensive studies on this topic, providing rich experience and data support that continue to drive the development of modern flow control technologies \cite{Tensi2002,feng2010circular,feng2012synthetic}. 

Several studies have investigated the influence of synthetic jets placement on flow control performance for flow past a square cylinder at low $Re$.
For instance, \citeauthor{chen2023deep} \cite{chen2023deep} investigated DRL-based AFC strategies to mitigate vortex-induced vibrations of a square cylinder at $Re = 100$. 
Their study evaluated three synthetic jets actuator positions along the sides of the square cylinder: the front, middle, and rear. The results demonstrated that placing the synthetic jets at the rear of the square cylinder achieved the best control performance with a vibration reduction of 96\%, compared to 86\% and 79\% for the front and middle positions, respectively.
Similarly, \citeauthor{jia2024jetsactuator} \cite{jia2024jetsactuator} investigated the effects of synthetic jets placement and width on flow control for a square cylinder using DRL algorithms.
At $Re = 100$, synthetic jets near the trailing edge achieved a 14.4\% drag reduction, while at $Re = 500$, synthetic jets near the leading edge reduced drag by 65.5\%.

\subsection{Critical issues to address in the present work}

Although several studies have investigated the impact of synthetic jet arrangements on flow control performance around square cylinders, the qualitative conclusions drawn at different $Re$ remain inconsistent. 
In recent research on flow control around square cylinders using deep reinforcement learning algorithms, \citeauthor{chen2023deep} \cite{chen2023deep} demonstrated that, at $Re = 100$, synthetic jets placed at the rear edges of the square cylinder outperformed those at the front edges in terms of control effectiveness. In contrast, \citeauthor{yan2023stabilizing} \cite{yan2023stabilizing} reported that, at higher $Re$ = 500, 1000, and 2000, jet control strategies positioned at the front edges of the square cylinder were more effective than those at the rear edges. 
This highlights the need for further exploration into the sensitivity of jet placement.

Furthermore, \citeauthor{jia2024jetsactuator} 
\cite{jia2024jetsactuator} found contradictory conclusions when comparing results at $Re = 100$ and $Re = 500$, systematically testing earlier findings. Their analysis revealed that placing synthetic jets near primary flow separation points achieves optimal control, with inconsistencies in prior conclusions attributed to changes in separation locations across $Re$. They also demonstrated that complete vortex shedding suppression requires jets positioned near the cylinder's trailing corners. 
While the effects of synthetic jet placement on square cylinders have been explored, its impact on circular cylinders remains unverified. Therefore, this study aims to address this gap in the existing research.

On the other hand, previous studies have provided a comprehensive understanding of active flow control at low $Re$ \cite{jiang2023reinforcement,jia2024optimal}. However, at medium-to-high $Re$, two critical challenges remain unresolved: insufficient drag reduction efficiency and the persistent instability of the controlled flow. These issues highlight the inherent complexity of achieving optimal control performance in such regimes.  
\citeauthor{Varela2022} \cite{Varela2022} demonstrated a 20\% drag reduction for flow around a cylinder at $Re = 1,000$, while \citeauthor{wang2024dynamic} \cite{wang2024dynamic} and \citeauthor{renApplying} \cite{renApplying} reported more substantial drag reduction of 30\% and 46\% for flow around a cylinder, respectively, at the same $Re$. 

Despite these advancements, the controlled flows exhibited significant fluctuations in both drag and lift coefficients, suggesting that the implemented control strategies failed to stabilize the flow effectively and fell short of achieving truly optimal performance.
Furthermore, \citeauthor{Varela2022} \cite{Varela2022} extended their work to flow around a cylinder at $Re = 2,000$, achieving a drag reduction of 18\%. However, the controlled flow is still very unstable and the drag reduction is limited. This further emphasizes the limitations of existing research. Therefore, this study aims to explore stable control performance at medium to high $Re$.

This study focuses on analyzing the sensitivity of flow control performance to variations in synthetic jet placement and achieving complete suppression of vortex shedding under complex dynamics at $Re$ of 500, 1,000, and 2,000.
In section \ref{sec:Methodology}, we provide an overview of the research methodology, including numerical simulations and the DRL framework. Section \ref{sec:Results} presents the training results, emphasizing the effectiveness of the proposed control strategy. Finally, section \ref{sec:Conclusions} summarizes the main findings and discusses the significance of the study.

\section{METHODOLOGY}\label{sec:Methodology}

\begin{figure*}[t]
    \centering
    \includegraphics[width=\textwidth]{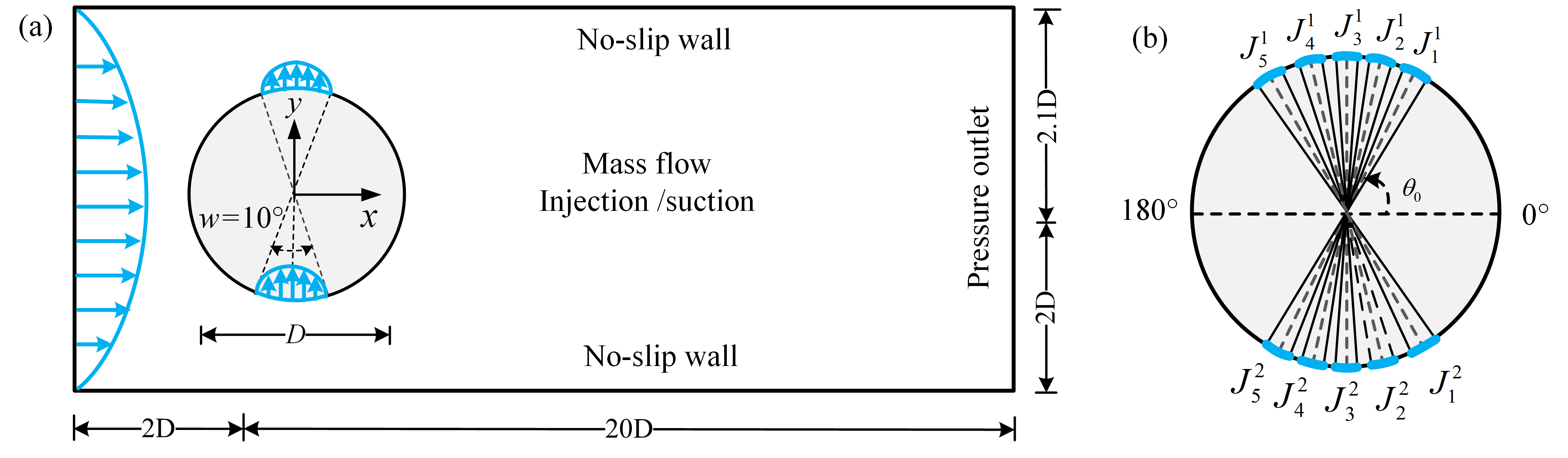}
    \caption{Schematic of the computational domain for the flow simulation. (a) Computational domain for flow around a circular cylinder with synthetic jets. (b) Arrangement of synthetic jets ($ J_i^1 $ and $ J_i^2 $) on the cylinder surface, with $ w = 10^\circ $ spacing and initial angle $\theta_0$.
    }
\label{fig:figure1}
\end{figure*}

Based on \citeauthor{jia2024optimal} \cite{jia2024optimal}, we further explore a DRL-based flow control framework with multiple control objectives, aiming to reduce drag, minimize lift, and suppress vortex shedding.
Central to this framework is the use of strategically positioned observation probes that provide insights to guide the DRL agent, which dynamically modulates the mass flow rate of synthetic jets while interacting with a computational fluid dynamics environment to refine its control policy. Our investigation focuses on examining the sensitivity of control performance to synthetic jet placement and the robustness of strategies under moderate-to-high Reynolds number regimes.

\subsection{Flow Configuration}\label{sec:Simulation}
We study the wake flow behind a two-dimensional cylinder in a confined domain, as shown in figure \ref{fig:figure1}. The configuration follows the classical benchmark by Sch\"afer \textit{et al}.\cite{Schafer1996}. A Cartesian coordinate system is used, with $x$ representing the streamwise direction and $y$ the wall-normal direction. All spatial dimensions are scaled by the cylinder diameter $D$. The computational domain is centered on the cylinder, extending $20D$ downstream and $2D$ upstream, with a total length of $22D$. In the wall-normal direction, the domain spans $2.1D$ above and $2D$ below the cylinder, giving a total width of $4.1D$. Similar to the benchmark by Sch"afer \textit{et al.} \cite{Schafer1996}, this asymmetric design is intended to help trigger vortex shedding.

The Navier-Stokes equations for an incompressible, viscous fluid within a domain $\Omega \subset \mathbb{R}^{nd}$ over a time interval $(0, T)$ constitute the fundamental framework for fluid dynamics analysis, expressed as follows:
\begin{subequations}
\begin{equation}
    \frac{\partial \mathbf{u}}{\partial t} + \mathbf{u} \cdot \nabla \mathbf{u} = -\nabla p + \frac{1}{Re} \nabla^2 \mathbf{u},  \quad \text{in} \quad \Omega \times (0, T),
\end{equation}
\begin{equation}
    \nabla \cdot \mathbf{u} = 0, \quad \text{in} \quad \Omega \times (0, T),
\end{equation}
\end{subequations}
where $\mathbf{u} = \mathbf{u}(\mathbf{x}, t)$ represents the fluid velocity field (with $\mathbf{x}$ as the spatial coordinates and $t$ as time), $p = p(\mathbf{x}, t)$ is the pressure field, and the Reynolds number is defined as 
$Re = \frac{\overline{U}L}{\nu}$ where $\overline{U}$ is the characteristic velocity at the inlet, $L$ is the characteristic length scale of the domain (equal to the cylinder diameter $D$ for flow around a cylinder), and $\nu$ is the kinematic viscosity of the fluid.

The flow around a circular cylinder in confined conditions follows the benchmark by ~\citeauthor{Schafer1996} \cite{Schafer1996}. 
At the inlet $\Gamma_\text{in}$, the inflow velocity along $x$-axis is prescribed by a parabolic velocity profile in the form,
\begin{equation}\label{equ:1velocity}
U_{\text{inlet}}(y)=U_m\frac{(H-2y)(H + 2y)}{H^2},
\end{equation}
where $U_m$ is the maximum velocity magnitude of the parabolic profile, and $H=4.1D$ represents the total height of the rectangular domain. The average inlet velocity $\overline{U}$, is related to the parabolic velocity profile $U_{\text{inlet}}(y)$ through the expression:
\begin{equation}\label{equ:mean velocity}
\overline{U}=\frac{1}{H} \int_{-H/2}^{H/2} U_{\text{inlet}}(y) d y=\frac{2}{3} U_m.
\end{equation}

The outflow boundary $ \Gamma_{\text{out}} $, allows fluid to exit the domain, utilizing a pressure outlet condition to ensure proper flow.
The synthetic jet boundaries $ \Gamma_{i} $ ($ i = 1, 2 $), employ a parabolic velocity profile for fluid injection and suction. 
To ensure mass flow conservation, the velocity magnitudes satisfy the condition $ V_{\Gamma_1} = -V_{\Gamma_2} $.  
The corresponding boundary conditions are mathematically expressed as:
\begin{align}
-\rho \mathbf{n} \cdot \mathbf{p} + Re^{-1} (\mathbf{n} \cdot \nabla \mathbf{u}) &= 0  \quad \text{on } \Gamma_{\text{out}}, \\
\mathbf{u} &= 0  \quad \text{on } \Gamma_{\text{w}}, \\
\mathbf{u} &= U_\text{inlet}  \quad \text{on } \Gamma_{\text{in}}, \\
\mathbf{u} &= f_{Q_i}  \quad \text{on } \Gamma_{i}, \quad i = 1, 2,
\end{align}
where $\mathbf{n}$ denotes the normal vector, and $f_{Q_i}$ represents the radial velocity distribution simulating the suction or injection of fluid by the jets. 
For a jet on the surface of a cylinder with diameter $D$, centered at $\alpha_0$, and with width $\omega$, the radial velocity profile is given by 
\begin{eqnarray}
u_{\text{jet}}(\alpha, Q_i) = \frac{\pi}{\omega D} Q_i \cos\left(\frac{\pi}{\omega} (\alpha - \alpha_0) \right),
\end{eqnarray}
where $\omega = 10^\circ$ and $Q_i$ $(i = 1, 2)$ is the mass flow rate that determines the jet's intensity.
The cosine function modulates the velocity, smoothly reducing it to zero at the boundaries. 

The placement of synthetic jets is a key aspect of this study. The jets are arranged in five distinct configurations, each defined by an azimuthal angle $\theta_0$ relative to the positive $x$-axis. Specifically, the first jet in each symmetric pair is positioned at $\theta_0 = 60^\circ, 75^\circ, 90^\circ, 105^\circ,$ and $120^\circ$, while the second jet is symmetrically mirrored across the $x$-axis. To systematically denote their positions, each jet is labeled as $J_i^j$, where the subscript $i$ ($i = 1, \dots, 5$) corresponds to the azimuthal angle, and the superscript $j$ ($j = 1, 2$) distinguishes between the first and second jet in the pair. This structured placement ensures balanced jet distribution across the selected angles, facilitating a comprehensive evaluation of their influence on flow control.

Numerical simulations in this study are conducted using the open-source \texttt{OpenFOAM} platform, which employs the finite volume method for domain discretization \cite{jasak2009}. 
The simulations are performed with the \texttt{pimpleFoam} solver, which integrates the Pressure-Implicit with Splitting of Operators (\texttt{PISO}) and Semi-Implicit Method for Pressure-Linked Equations (\texttt{SIMPLE}) algorithms to ensure numerical stability and efficiency \cite{SIMPLE}.
To validate the accuracy of our baseline flow simulations, we compare the computed drag and lift coefficients at $Re = 500$, 1,000, and 2,000 with results from previous studies, as summarized in table \ref{tab:table2}. These studies follow the benchmark established by \citeauthor{Schafer1996} \cite{Schafer1996}, providing a reliable reference for assessing the fidelity of our numerical approach.

\begin{table}[htb!]
\centering
\begin{threeparttable}
\caption{Comparison of numerical simulations with previous studies.}
\begin{tabularx}{\textwidth}{
  >{\centering\arraybackslash}p{0.15\textwidth}
  >{\centering\arraybackslash}p{0.2\textwidth}
  >{\centering\arraybackslash}p{0.15\textwidth}
  >{\centering\arraybackslash}p{0.15\textwidth}
  >{\centering\arraybackslash}p{0.15\textwidth}
}
\toprule
\toprule
$Re$ & Reference & $C_{D,mean}$ & $C_{D,std}$ & $C_{L,std}$  \\
\midrule
500 & Our work & 3.15 & 0.273 & 2.12  \\ 
500 & \citeauthor{wang2024dynamic} \cite{wang2024dynamic} & 3.20 & 0.283 & 2.17  \\ 
1000 & Our work & 3.42 & 0.438 & 2.70  \\
1000 & \citeauthor{renApplying} \cite{renApplying} & 3.44 & - & -  \\
1000 & \citeauthor{wang2024dynamic} \cite{wang2024dynamic} & 3.48 & 0.455 & 2.76  \\ 
2000 & Our work & 3.74 & 0.512 & 3.16  \\
2000 & \citeauthor{Varela2022} \cite{Varela2022} & 3.39 & - & - \\ 
\bottomrule
\bottomrule
\end{tabularx}
\begin{tablenotes}
\footnotesize
\item Note: $C_{D,std}$ and $C_{L,std}$ are the temporal standard deviations over the last 100 non-dimensional time units in statistically steady state.
\end{tablenotes}
\label{tab:table2}
\end{threeparttable}
\end{table}

\begin{figure*}
    \centering
    \includegraphics[width=\textwidth]{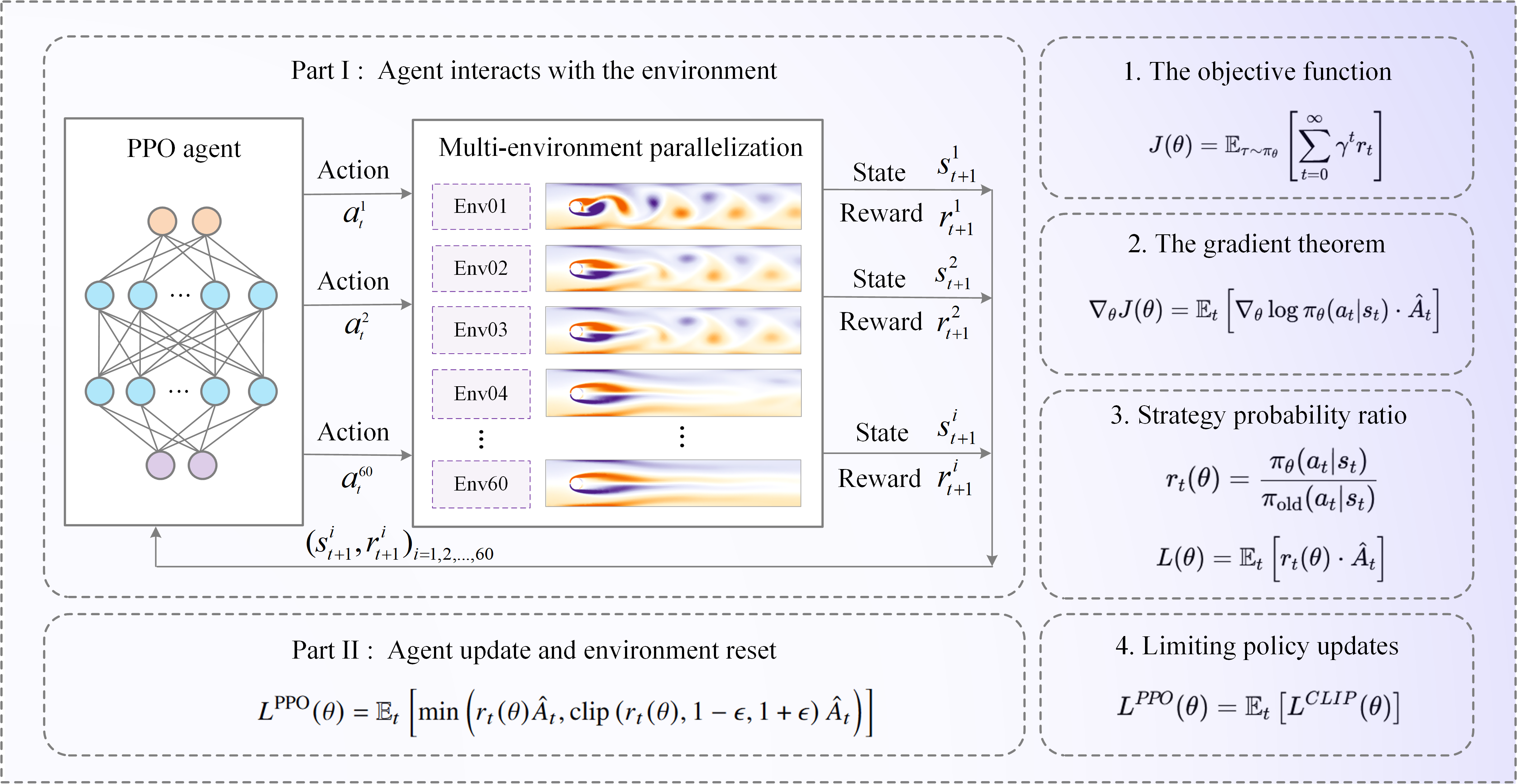}
    \caption{Workflow of proximal policy optimization for active flow control: agent-environment interaction and policy update framework and key equations of PPO algorithm.}
\label{fig:figure2}
\end{figure*}

\subsection{Deep reinforcement learning}\label{sec:reinforcement}

Reinforcement learning (RL) is a machine learning framework where an agent learns to make decisions by interacting with an environment to maximize cumulative rewards \cite{van2012reinforcement}. Mathematically, RL is modeled as a Markov Decision Process (MDP) defined by $(S, A, P, R, \gamma)$, where $S$ and $A$ represent the state and action spaces, $P(s'|s, a)$ is the state transition probability, $R(s, a)$ is the reward function, and $\gamma$ is the discount factor \cite{puterman1990markov}. The agent generates trajectories by selecting actions $a_t$ based on a policy $\pi(a|s)$, transitioning to new states $s_{t+1}$ according to $P(s_{t+1}|s_t, a_t)$, and receiving rewards $r_t$ \cite{franccois}. The trajectory is formally expressed as a sequence of states, actions, and rewards: 
\begin{equation}
\tau = \{(s_0, a_0, r_0), (s_1, a_1, r_1), \dots, (s_T, a_T, r_T)\},
\end{equation}
where $T$ denotes the trajectory length, determined by either a predefined horizon or a termination condition imposed by the environment.
The aims to optimize the policy $\pi^*$ that maximizes the cumulative reward over an episode, defined as:  
\begin{equation}
R(\tau) = \sum_{t=1}^n \gamma^t r_t,
\end{equation} 
where $\gamma$ is the discount factor balancing short-term and long-term rewards. In DRL, this optimization translates to determining the optimal parameters $\theta^*$ of the artificial neural network.

The state-value function $V^\pi(s)$ evaluates the expected cumulative reward starting from state $s$ under policy $\pi$, providing a measure of the overall value of a state. In contrast, the action-value function $Q^\pi(s, a)$ evaluates the expected cumulative reward for taking action $a$ in state $s$ under policy $\pi$, enabling comparison of different actions. These functions are formally defined as:
\begin{equation}
\begin{aligned}
V^\pi(s) &= \mathbb{E} \left[ \sum_{k=0}^\infty \gamma^k r_{t+k} \,\middle|\, s_t = s \right], \\
Q^\pi(s, a) &= \mathbb{E} \left[ \sum_{k=0}^\infty \gamma^k r_{t+k} \,\middle|\, s_t = s, a_t = a \right]. \\
\end{aligned}
\end{equation}

To connect these two functions, the advantage function $A^\pi(s, a)$ is introduced, which quantifies the relative benefit of taking action $a$ in state $s$ compared to the policy's default behavior. It is defined as the difference between the action-value function and the state-value function: $A^\pi(s, a) = Q^\pi(s, a) - V^\pi(s)$. 
Policy gradient methods aim to directly optimize the policy $\pi_{\theta}(a|s)$, parameterized by $\theta$, to maximize the expected return. The objective function for the expected return is defined as:
\begin{equation}
J(\pi_{\theta}) = \mathbb{E}_{\pi_{\theta}} \left[ \sum_{t=0}^{\infty} \gamma^t R_t \right].
\end{equation}

Proximal Policy Optimization (PPO) is a gradient-based reinforcement learning algorithm designed to improve the stability of policy updates while optimizing the expected return \cite{schulman2017}. It builds upon traditional policy gradient methods by introducing a clipping mechanism to prevent overly large policy updates, thereby enhancing training robustness.  
The objective function for PPO is defined as :  
\begin{equation}
L^{\text{CLIP}}(\theta) = \mathbb{E}_t \left[ \min \left( r_t(\theta) A_t, \text{clip}(r_t(\theta), 1 - \epsilon, 1 + \epsilon) A_t \right) \right],
\end{equation}
where $r_t(\theta) = \frac{\pi_{\theta}(a_t|s_t)}{\pi_{\theta_{\text{old}}}(a_t|s_t)}$ is the probability ratio between the new and old policies, representing how the updated policy deviates from the previous one. $A_t$ is the advantage function, which quantifies the relative benefit of an action compared to the average performance of the policy. $\epsilon$ is a hyperparameter that defines the clipping range, restricting the magnitude of policy updates to maintain stability.

\subsection{Framework of Active Flow Control with DRL}

\begin{figure*}
    \centering
    \includegraphics[width=\textwidth]{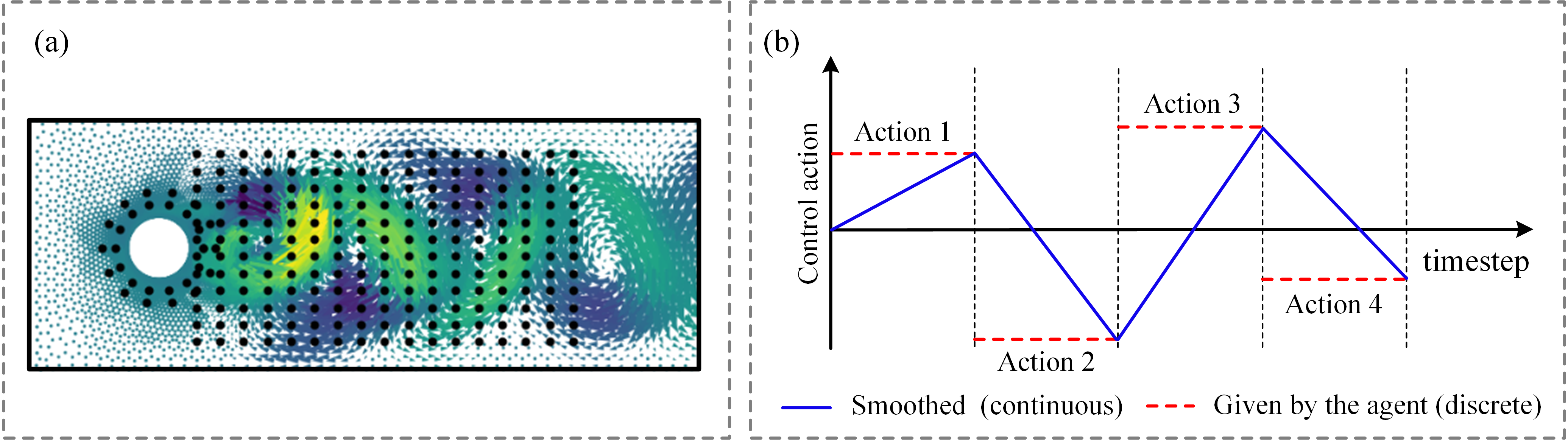}
    \caption{(a) Observation probe distribution. (b) Smooth interpolation of control actions for active flow control.}
\label{fig:figure3}
\end{figure*}

We integrate computational fluid dynamics simulations with DRL-based control strategies to establish a closed-loop system for real-time decision-making. The PPO-based active flow control framework is illustrated in figure \ref{fig:figure2}.
Part I presents the agent-environment interaction, where the PPO agent generates actions for multiple environments simulating wake flows, receiving state updates and rewards. Part II describes the policy update process, incorporating key PPO equations. This parallel setup enables efficient learning of optimal flow control strategies.

\begin{itemize}
    \item \textit{Interaction process.} At each step, the PPO agent observes the current state $ s_t $ and selects an action $ a_t $ based on its policy. This action is applied to the environment, which updates the flow dynamics and transitions to a new state $ s_{t+1} $. The effectiveness of the action is quantified through a reward $ r_t $, guiding the learning process. The agent continuously collects trajectory data $ (s_t, a_t, r_t, s_{t+1}) $ and, after accumulating sufficient experience, updates its policy using PPO. 
    
    \item \textit{Policy updates.} 
    The policy $ \pi_{\theta}(a_t | s_t) $ is represented by a neural network, which maps observed states $ s_t $ to action probabilities. During training, PPO optimizes this neural network using a clipped surrogate objective function $ L^{CLIP}(\theta) $, preventing drastic policy updates that could destabilize learning. The value function $ V_{\phi}(s_t) $, also approximated by a neural network, estimates future rewards to enhance decision-making. 
    The policy network and value network each consist of two hidden layers, with 512 neurons per layer, and utilize the $\tanh$ activation function. The learning rate is set to 0.001, the batch size is 60, and the discount factor is 0.97.

    \item \textit{Design of controlled actions.} 
    The smoothing function mitigates abrupt variations in control inputs, emulating actuator inertia and improving numerical stability. To ensure continuity and prevent discontinuities in pressure and velocity fields, particularly for incompressible solvers, the jet velocity is updated at the end of each actuation period as:  
    \begin{equation}
        Q_{\Gamma_1,T_i} = Q_{\Gamma_1,T_{i-1}} + \beta (a - Q_{\Gamma_1,T_{i-1}}),
    \end{equation}
    where $Q_{\Gamma_1,T_i}$ is the jet velocity at the end of the $i$-th period, $a$ is the agent's action, $Q_{\Gamma_1,T_{i-1}}$ is the velocity at the previous period, and $\beta = 0.1$ controls the transition rate \cite{rabault2019artificial,tangRobust2020}.
 
    \item \textit{Observation of environmental state.} A total of 229 sensor probes are strategically placed around the cylinder and in the wake region, with a higher density concentrated in areas of significant flow fluctuations. These probes are designed to capture detailed observations of the flow dynamics, as shown in figure \ref{fig:figure3}. They measure key flow characteristics, such as pressure, velocity, and vorticity distributions, which are then relayed to the agent for decision-making. The design of the probe distribution is informed by extensive studies from previous research, ensuring its effectiveness in monitoring the flow behavior \cite{parisRobust, liReinforcementlearning}.
   
    \item \textit{Definition of reward mechanism.} 
    To maximize drag reduction, the instantaneous reward function is defined as:
    \begin{eqnarray}
    r_{T_i} = C_{D,0} - \left(C_D\right)_{T_i} - \omega \left| \left(C_L\right)_{T_i} \right|,
    \end{eqnarray}
    where the drag coefficient ($C_D$), lift coefficient ($C_L$), and baseline drag coefficient ($C_{D,0}$) are calculated as the average values over the driving cycle $T_i$.
    The reward function is designed to balance drag reduction and suppression of lift fluctuations. 
    It incentivizes drag reduction by minimizing the difference between the baseline and instantaneous drag coefficients, while penalizing large lift oscillations through a term proportional to the absolute lift coefficient, with the weighting factor $\omega$ controlling the trade-off \cite{rabault2019artificial,jia2024optimal}. 

    \item \textit{Other training related details.} 
    For the DRL implementation, the agent is developed using the \textit{Tensorforce} platform with the PPO algorithm \cite{2017tensorforce}, while the environment is constructed via the \texttt{Gym} interface \cite{brockman2016openai}. 
    To accelerate convergence, 60 parallel CFD environments are employed, which efficiently generate training data for policy updates based on the method proposed by \citeauthor{jia2024optimal} \cite{jia2024optimal}.
    Each training session utilizes all 64 CPU cores, with 60 cores assigned to parallel CFD environments that generate 60 trajectories per batch in approximately 8 minutes. A total of 3000 episodes are conducted to ensure stable convergence, resulting in a total training time of about 6.7 hours.
\end{itemize}

\section{RESULTS}\label{sec:Results}

In this section, we conducted a comparative analysis of the control performance of five synthetic jet configurations at $ Re = 500 $ and $ Re = 1,000 $. Additionally, we thoroughly examined the flow control performance at $ Re = 500 $, $ Re = 1,000 $, and $ Re = 2,000 $, analyzing the effectiveness of the control strategies and their underlying mechanisms based on physical phenomena.

\subsection{Impact of synthetic jet configurations on flow control performance at Re = 500.}\label{sec:sec3}

\begin{figure*}[hbt!]
    \centering
    \includegraphics[width=\textwidth]{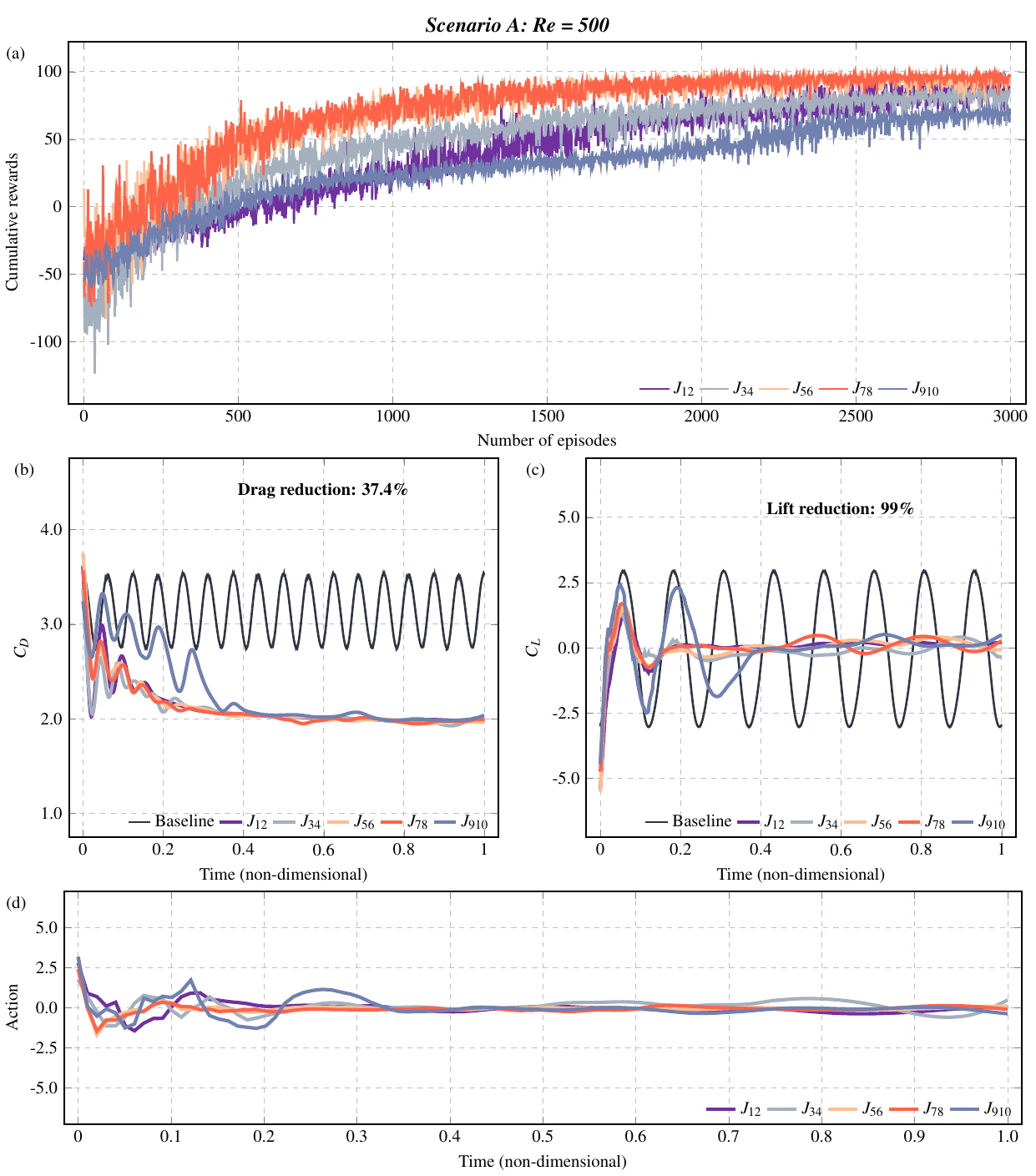}
    \caption{Performance analysis of synthetic jet configurations at $Re$ = 500. (a) Cumulative rewards versus number of episodes for different synthetic jet configurations. (b) Drag coefficient evolution over non-dimensional time for the baseline and synthetic jet configurations. (c) Lift coefficient evolution over non-dimensional time for the baseline and synthetic jet configurations. (d) Temporal evolution of actions for different synthetic jet configurations over non-dimensional time.}
\label{fig:figure4}
\end{figure*}

For flow around a cylinder at $ Re = 500 $, we evaluate the impact of five synthetic jet configurations on control performance. 
We first evaluate the stability and convergence of the reward function, which are key metrics for DRL. We then quantify flow control performance through drag reduction rates and the suppression of lift fluctuations, providing a comprehensive understanding of the achieved benefits.
To further assess the efficiency of the control strategies, we analyze the evolution of jet control actions to estimate the energy expenditure associated with external flow manipulation. The influence of synthetic jet placement on flow control effectiveness is specifically examined at $ Re = 500 $, with the results presented in figure \ref{fig:figure4}.

The evolution of the cumulative reward function during DRL training over the number of episodes is shown in figure \ref{fig:figure4} (a) for the positions of the synthetic jets at $ J_{12} $, $ J_{34} $, $ J_{56} $, $ J_{78} $, and $ J_{910} $.
In the case of the five synthetic jet placement configurations, the cumulative reward function exhibits a stepwise growth trend as the number of training episodes increases. This evolution of the reward function is consistent with the observations reported by \citeauthor{rabaultAccelerating} \cite{rabaultAccelerating}, confirming the alignment of the observed trend with their findings.
Initially, the reward function is negative, indicating that the agent has yet to develop a control strategy that fulfills the desired control objectives. However, as the agent refines its policy through iterative learning, the reward steadily improves, eventually converging to its maximum value, signifying the successful acquisition of an effective control strategy.

When the synthetic jets are positioned at $ J_{56} $ and $ J_{78} $, the system exhibits the fastest convergence, most stable results, and highest reward function values, demonstrating that these configurations are particularly effective for flow control. In contrast, when the jets are positioned at $ J_{910} $, the system shows the slowest convergence, poorest stability, and lowest reward function values, indicating that this configuration is the least effective for flow control. The performance of the configurations at $ J_{12} $ and $ J_{34} $ falls between these extremes, with moderate convergence and stability in the DRL training process.

In figures \ref{fig:figure4} (b) and (c), we provide a detailed comparison of the flow control performance corresponding to various synthetic jet placements, including five configurations—$J_{12}$, $J_{34}$, $J_{56}$, $J_{78}$, and $J_{910}$—with the baseline flow serving as a reference. In the baseline flow, both $C_D$ and $C_L$ exhibit strong periodic fluctuations, characteristic of typical unsteady wake dynamics. Under the five synthetic jet configurations, both the drag coefficient and lift coefficient of the controlled flow are effectively reduced, and the characteristics of periodic oscillations are completely eliminated.
Among these configurations, we observe that for the synthetic jet placed at $J_{910}$, the fluctuations in the lift coefficient, drag coefficient, and the control action evolution curve over time are significantly higher compared to the other configurations. This indicates that the control performance of the synthetic jet at $J_{910}$ is the most unstable.

To gain further insights into the energy consumption and stability of each jet configuration, figure \ref{fig:figure4} (d) illustrates the evolution of the flow control actions over time, which represents the control strategy.
The control strategies resulting from DRL training under various jet configurations indicate that synthetic jets placed at locations $ J_{56} $ and $ J_{78} $ produce control signals that are smoother and more stable.
This indicates that flow control is performed with lower energy consumption and higher efficiency in both jet configurations.
In contrast, the control actions for the synthetic jet pairs at $J_{12}$ and $J_{910}$ show more unstable performance, reflecting that they are less energy efficient in flow control than the other locations.

\begin{table*}
\centering
\caption{Flow control performance metrics for different jet configurations at $Re = 500$.}
\begin{tabularx}{\textwidth}{
  >{\centering\arraybackslash}p{0.05\textwidth}
  >{\centering\arraybackslash}p{0.05\textwidth}
  >{\centering\arraybackslash}p{0.1\textwidth}
  >{\centering\arraybackslash}p{0.2\textwidth}
  >{\centering\arraybackslash}p{0.1\textwidth}
  >{\centering\arraybackslash}p{0.1\textwidth}
  >{\centering\arraybackslash}p{0.1\textwidth}
  >{\centering\arraybackslash}p{0.1\textwidth}  
}
\toprule
\toprule
$Re$ & Jet & $C_\text{D, mean}$ & Drag reduction (\%) & $C_\text{D, std}$ & $C_\text{L, std}$ & $Q_\text{mean}$ & $Q_\text{std}$ \\
\midrule
500  & $J_{12}$  & 1.986 & 37.0  & 0.028 & 0.247 & -0.185 & 0.130 \\
500  & $J_{34}$  & 1.979 & 37.2  & 0.024 & 0.251 &  0.100 & 0.392 \\
500  & $J_{56}$  & 1.972 & 37.4  & 0.015 & 0.148 & -0.044 & 0.072 \\
500  & $J_{78}$  & 1.979 & 37.2  & 0.019 & 0.208 & -0.031 & 0.091 \\
500  & $J_{910}$ & 2.001 & 36.5 & 0.026 & 0.262 & -0.110 & 0.120 \\
\bottomrule
\bottomrule
\end{tabularx}
\label{tab:table3}
\end{table*}

We quantify the control performance in table \ref{tab:table3} to more clearly illustrate the sensitivity of control performance to different synthetic jet placements. The drag reduction rate is quantified by the decrease in the drag coefficient of the controlled flow relative to the baseline flow. The stability of the controlled flow is characterized by the standard deviation of the drag coefficient and lift coefficient time series. Additionally, the required external energy to maintain the flow control effect is quantified by the mean and standard deviation of the synthetic jet control action during the second half of the control period.

As observed from table \ref{tab:table3}, the drag reduction achieved when the synthetic jets are positioned at $ J_{910} $ is 36.5\%, which is notably lower than the drag reduction observed for other configurations. The optimal drag reduction rates of 37.4\% and 37.2\% are achieved when the jets are placed at $ J_{56} $ and $ J_{78} $, respectively. Furthermore, the standard deviations of both the drag and lift coefficients are considerably smaller for these configurations, suggesting superior stability of the controlled flow compared to other placements. The mean values and standard deviations of the synthetic jet control action at these positions are an order of magnitude lower than those observed at other locations, indicating that these configurations minimize the external energy required for effective flow control.

\begin{figure*}[hbt!]
    \centering
    \includegraphics[width=\textwidth]{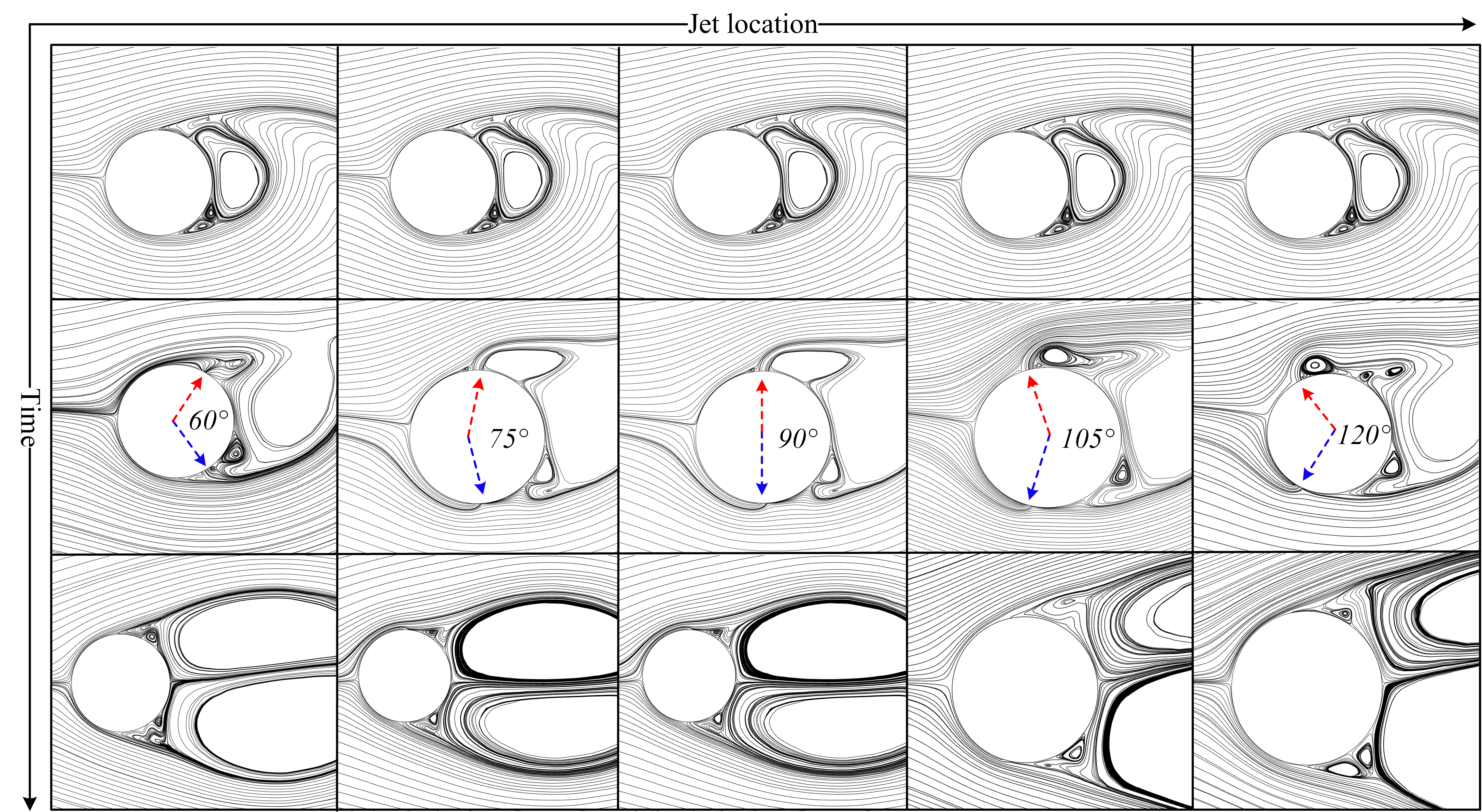}
    \caption{The evolution of flow structures for different synthetic jet placements at angular positions $\theta = 60^\circ$, $\theta = 75^\circ$, $\theta = 90^\circ$, $\theta = 105^\circ$, and $\theta = 120^\circ$.
    }
\label{fig:figure5}
\end{figure*}

To further investigate the control mechanisms of flow control using synthetic jets, we present the evolution of streamlines around the cylinder in figure \ref{fig:figure5}, where the synthetic jets are placed at positions of $\theta = 60^\circ$, $\theta = 75^\circ$, $\theta = 90^\circ$, $\theta = 105^\circ$, and $\theta = 120^\circ$. For clarity, we specifically highlight three representative moments: before control, during the activation of the synthetic jets, and at the end of control, to illustrate the evolution of the streamlines around the cylinder at these key instances. Each column represents a different jet configuration. The first row presents the baseline flow without control. The second row shows the flow state when the synthetic jet actuation is activated, with one jet blowing and the other drawing in. The third row depicts the flow structure after the control is applied.

First, synthetic jets are inherently zero-mass-flux devices, meaning they do not inject or expel any net mass into or from the fluid system. Instead, they generate flow disturbances by periodically altering the flow characteristics through blowing and suction.
As shown in figure \ref{fig:figure5}, under five different jet placement configurations, we observe closed recirculation zones either near the jet or above the flow structures around the cylinder. The formation of these recirculation zones enables synthetic jets to exert significant global effects on scales much larger than the characteristic length scale of the jet itself, thereby influencing the overall lift and drag on the cylinder.
This phenomenon is consistent with the findings of \citeauthor{amitay2012modification, Tensi2002}\cite{amitay2012modification, Tensi2002}, who investigated the mechanisms of flow control using synthetic jets, further validating the important role of synthetic jets in flow control.

We conducted a comparative analysis of the effects of synthetic jets positioned at different locations on the flow lines around a cylindrical baseline flow. Our findings reveal that as the position of the synthetic jet moves from the rear side of the cylinder upstream, the location, number, and size of the recirculation areas formed around the cylinder gradually change.
Specifically, when the synthetic jet is positioned at a 60° angle, a closed recirculation region forms only near the jet. At a 105° angle, a closed recirculation area appears both downstream of the jet and on the rear side of the cylinder. 
Furthermore, at a 120° angle, multiple small vortices are observed on the rear side of the cylinder, indicating a more pronounced trend.
These results demonstrate that placing synthetic jets on a blunt body surface to create recirculation regions can significantly alter the flow characteristics. However, the degree of modification to the flow structure around the cylinder varies with the position of the synthetic jet.
\citeauthor{amitay2012modification} utilized smoke visualization to observe synthetic jets at angles ranging from 60° to 150° and analyzed the control mechanisms at different positions on the cylinder's surface \cite{amitay2012modification}. 
Our observations align closely with their findings, further validating the significant impact of synthetic jets on flow structures.

Additionally, we observed that after the control period, the vortices formed near the rear of the cylinder stabilized, and a nearly symmetrical flow separation bubble developed on the leeward side of the cylinder. The wake of the cylinder became fully stabilized, with no significant vortex shedding, resulting in a streamlined wake structure. This transition demonstrates the effectiveness of synthetic jets in controlling the flow and enhancing wake stability.

\subsection{Impact of synthetic jet configurations on flow control performance at Re = 1000. }

\begin{figure*}[hbt!]
    \centering
    \includegraphics[width=\textwidth]{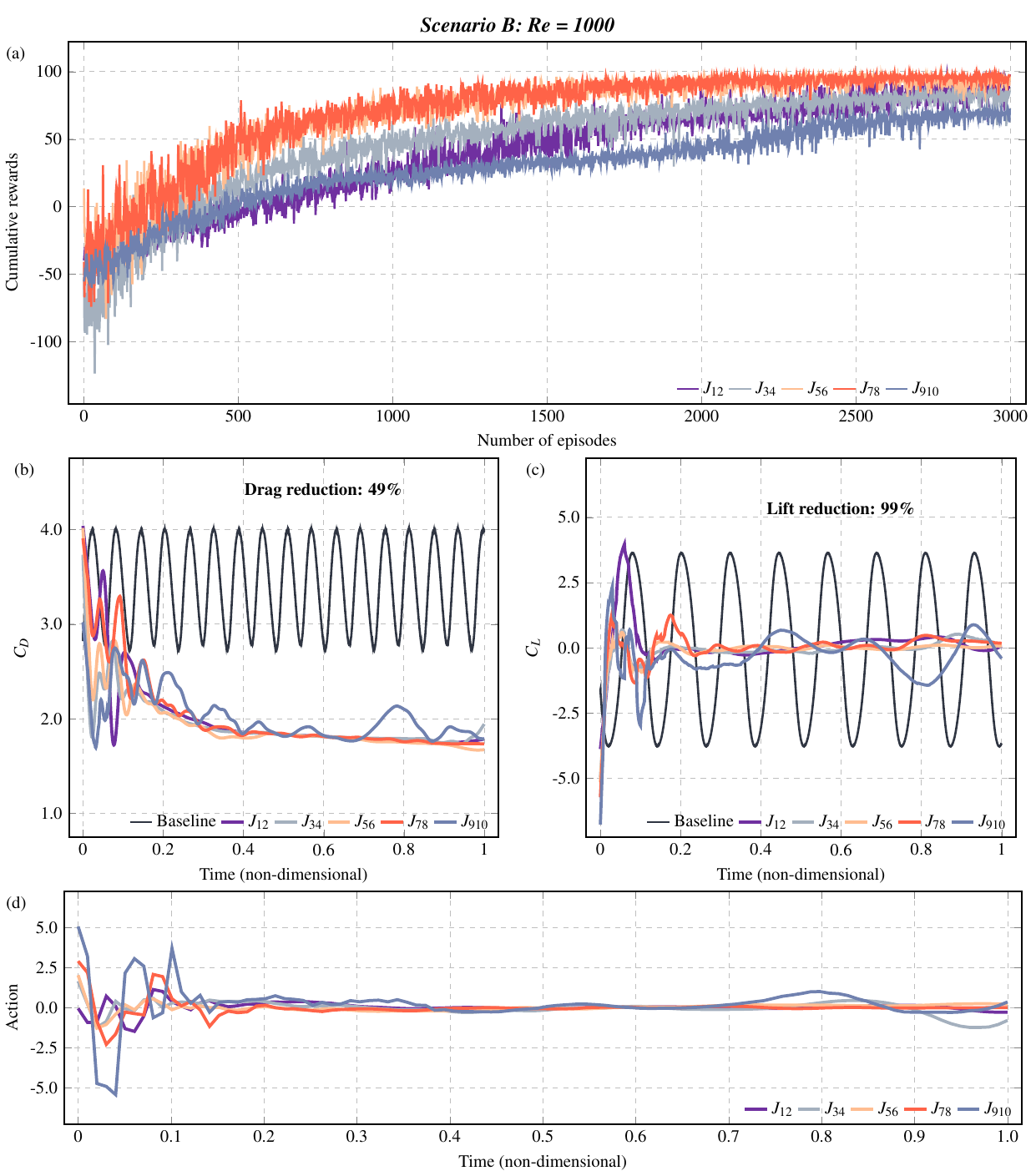}
    \caption{Performance analysis of synthetic jet configurations at $Re$ = 1000. 
    (a) Cumulative reward as a function of the number of episodes for different synthetic jet configurations.
    (b) Time history of the drag coefficient of a cylinder for the baseline flow and five jet configurations.
    (c) Time history of the lift coefficient of a cylinder for the baseline flow and five jet configurations. 
    (d) Evolution of the control strategy over time for five jet configurations.
    }
\label{fig:figure6}
\end{figure*}

In figure \ref{fig:figure6} (a), the development trend of the cumulative reward for various synthetic jet configurations at $Re = 1,000$ is largely consistent with the pattern observed at $Re = 500$.
The results show that the cumulative rewards for all configurations show a consistent upward trend, indicating that the control strategy gradually achieves the control goal.
The gradually decreasing fluctuations in the reward function value under each synthetic jet configuration reflect the convergence of DRL training and highlight the robustness of the reinforcement learning framework.
It is noteworthy that the cumulative reward value when the synthetic jets are positioned at $ J_{56} $ outperforms all other configurations. The configurations at $ J_{34} $ and $ J_{78} $ also exhibit commendable performance, but their cumulative rewards are still slightly lower than that of $ J_{56} $. In contrast, the reward function convergence is slowest for the configurations at $ J_{12} $ and $ J_{910} $, indicating suboptimal performance for these placements.

In figures \ref{fig:figure6} (b) and (c), the control effects on the drag coefficient and lift coefficient for various synthetic jet positions at $ Re = 1,000 $ are presented. The baseline flow exhibits strong periodic oscillations in both the drag coefficient and lift coefficient.
In contrast, under controlled conditions, all configurations lead to a marked reduction in the drag coefficient, completely suppressing its periodic fluctuations. At the same time, substantial suppression of the lift coefficient is also achieved. 
Importantly, when the synthetic jets are positioned at $ J_{910} $, the fluctuations in both the drag and lift coefficients are noticeably larger compared to other jet configurations. 
This trend is particularly pronounced at $ Re = 1,000 $ relative to $ Re = 500 $.

The temporal evolution of the actuation signals for different synthetic jet configurations is shown as a function of non-dimensional time, as illustrated in figure~\ref{fig:figure6}(d).
Each jet configuration exhibits pronounced fluctuations in actuation during the initial transient phase, reflecting the agent’s active modulation of mass flow rates as it seeks to establish effective and stable control.
After the initial fluctuations, the control actions for most jet configurations stabilize, with values gradually converging to a more constant range, particularly for the $ J_{56} $ and $ J_{78} $ configurations. These configurations seem to reach a steady state more quickly, reflecting the agent's ability to effectively adjust the jets for stable flow control.
In contrast, the $ J_{910} $ configuration exhibits larger fluctuations in control actions, especially during the early stages of the simulation. This suggests that the agent encounters more difficulty with this configuration, likely due to less optimal placement, which results in poorer control performance. In comparison, the control action curves for $ J_{56} $ and $ J_{78} $ are smoother and more stable, indicating these configurations offer better control performance.

\begin{table*}[t]
\centering
\caption{Flow control performance metrics for different jet configurations at $Re = 1,000$.}
\begin{tabularx}{\textwidth}{
  >{\centering\arraybackslash}p{0.05\textwidth}
  >{\centering\arraybackslash}p{0.05\textwidth}
  >{\centering\arraybackslash}p{0.1\textwidth}
  >{\centering\arraybackslash}p{0.2\textwidth}
  >{\centering\arraybackslash}p{0.1\textwidth}
  >{\centering\arraybackslash}p{0.1\textwidth}
  >{\centering\arraybackslash}p{0.1\textwidth}
  >{\centering\arraybackslash}p{0.1\textwidth}  
}
\toprule
\toprule
$Re$ & Jet & $C_\text{D, mean}$ & Drag reduction & $C_\text{D, std}$ & $C_\text{L, std}$ & $Q_\text{mean}$ & $Q_\text{std}$ \\
\midrule
1000 & $J_{12}$  & 1.764 & 48.3 \% & 0.016 & 0.176 & -0.004 & 0.124 \\
1000 & $J_{34}$  & 1.787 & 47.6 \% & 0.032 & 0.224 & -0.184 & 0.554 \\
1000 & $J_{56}$  & 1.737 & 49.1 \% & 0.034 & 0.062 &  0.168 & 0.043 \\
1000 & $J_{78}$  & 1.762 & 48.4 \% & 0.024 & 0.059 &  0.007 & 0.040 \\
1000 & $J_{910}$ & 1.904 & 44.2 \% & 0.116 & 0.748 &  0.255 & 0.436 \\
\bottomrule
\bottomrule
\end{tabularx}
\label{tab:table4}
\end{table*}

At $ Re = 1,000 $, the flow control performance metrics for various jet configurations are presented in table \ref{tab:table4}. From a drag reduction perspective, the synthetic jet positioned at $ J_{56} $ achieved a maximum drag reduction rate of 49.1\%, while the configuration at $ J_{910} $ attained a minimum drag reduction rate of 44.2\%.
Analyzing flow stability, the standard deviations of the drag coefficient and lift coefficient are an order of magnitude larger for the synthetic jet at $ J_{910} $ compared to other jet positions, indicating a more unstable flow state. Conversely, when the synthetic jet is located at $ J_{56} $ or $ J_{78} $, the standard deviations of the drag and lift coefficients reach their lowest levels, suggesting these configurations offer better flow stability.

We analyzed the average and standard deviation of jet actuations during the latter half of the control cycle, which reflects the energy consumption required for sustaining flow control in the mid-to-late stages. The training results indicate that the synthetic jets at $J_{56}$ and $J_{78}$ exhibit the lowest average and standard deviation of actions, demonstrating both the stability of these configurations and their efficiency in external energy consumption.  
Similar to the case at a $Re$ of 500, the control strategy with the synthetic jet positioned at $J_{910}$ exhibits the highest instability during the early stages of control. Moreover, in the mid-to-late stages, maintaining control performance at this position requires the highest and most fluctuating energy consumption.  
In summary, the $J_{56}$ configuration excels in both drag reduction and flow stability, whereas $J_{910}$ results in lower performance and higher energy consumption. These DRL-based training results highlight the critical importance of selecting the optimal jet position to enhance flow control performance in practical applications.

Based on the flow control performance comparison presented in figure \ref{fig:figure6} and table \ref{tab:table4}, the synthetic jets in the $J_{56}$ and $J_{78}$ configurations exhibit superior control effectiveness compared to other jet arrangements. Specifically, the $J_{56}$ and $J_{78}$ configurations are positioned near the natural separation points of the boundary layer on either side of the cylinder. This configuration enables more efficient interaction between the synthetic jets and the separated shear layers. The induced perturbations facilitate momentum transfer toward the wall region, thereby energizing the boundary layer and delaying flow separation. As a result, vortex shedding is more effectively suppressed, and the wake is significantly stabilized. This mechanism is consistent with previous studies on the use of synthetic jets for separation control \cite{Glezer2002,Syntheticsteven}.

\begin{figure}[hbt!]
    \centering
    \includegraphics[width=0.47\textwidth]{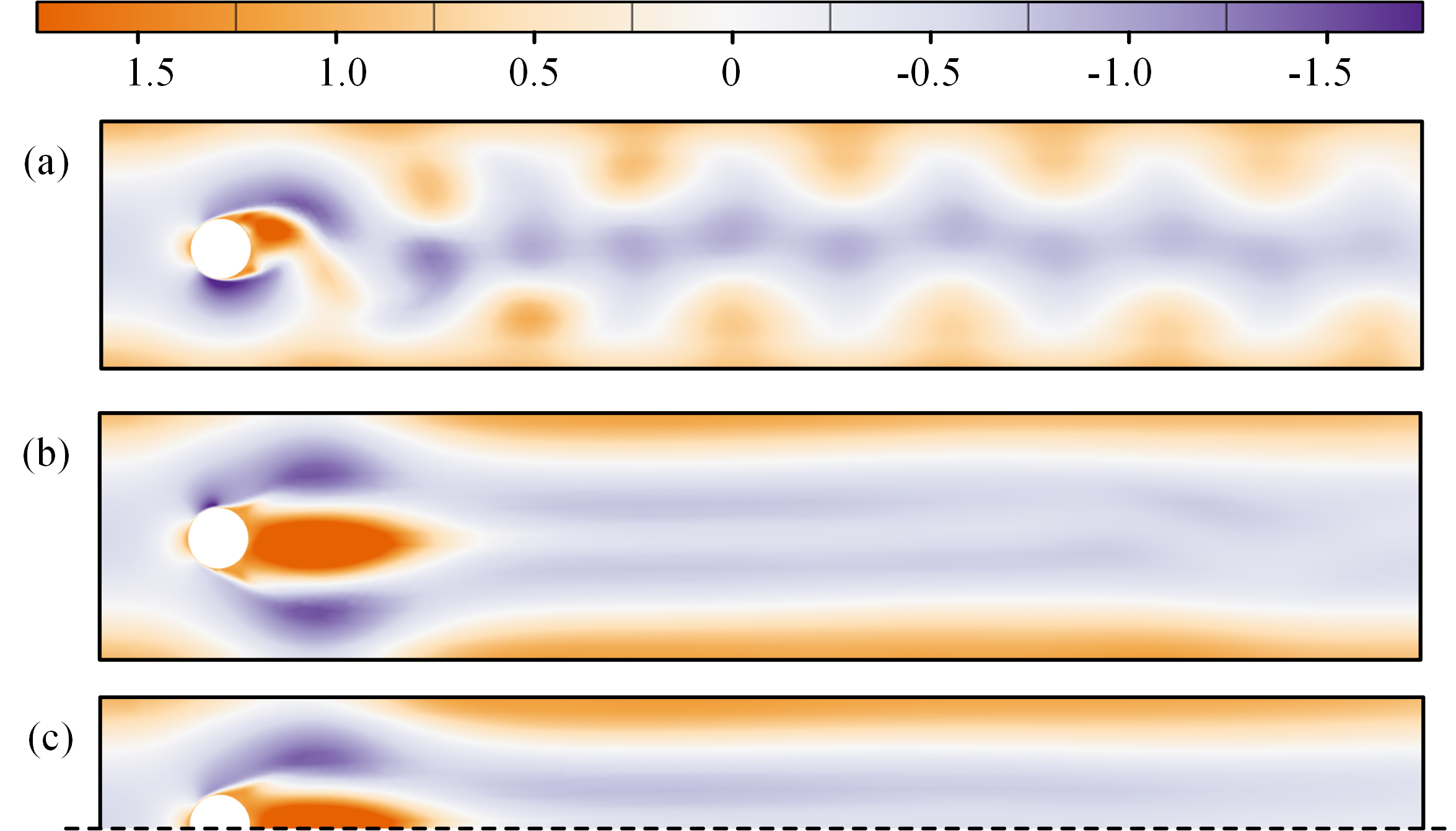}
    \caption{(a) The uncontrolled flow, (b) the controlled flow, and (c) the reference boundaries following Bergman et al.\cite{bergmann}.}
\label{fig:figure7}
\end{figure}

To assess the efficacy of the control strategy, we adopt a comparative framework inspired by the methodology of \citeauthor{bergmann}, which is also employed by Jia et al. \cite{jia2024jetsactuator} in the evaluation of control performance. 
\citeauthor{bergmann} proposes that the total drag acting on a circular cylinder consists of two principal components: one associated with the symmetric base flow around the body, and the other originating from vortex shedding at the cylinder surface \cite{bergmann}.
To estimate the contribution of the symmetric base flow to the total drag, we performed simulations of flow past a semi-cylinder at Reynolds number $Re = 1000$, imposing symmetry boundary conditions along the channel centerline. 
The resulting flow field is shown in figure \ref{fig:figure7}. In this case, the time-averaged drag coefficient on the semi-cylinder is $C_D = 0.927$.
Based on this result, and following the reasoning of \citeauthor{bergmann}, the drag coefficient of the full cylinder under complete suppression of vortex shedding is expected to asymptotically approach or fall below $C_D = 1.854$, corresponding to a drag reduction rate of approximately 47\% or higher \cite{bergmann}.

In the study by \citeauthor{renApplying}, the control strategy trained using DRL achieved a drag reduction rate of 34\% \cite{renApplying}. Correspondingly, the control strategy did not fully suppress vortex shedding from the cylinder. \citeauthor{renApplying} also candidly speculated that further improvements could be realized by employing more refined actuation \cite{renApplying}.
In stark contrast, the control strategy generated through our DRL framework demonstrates a significant advancement, achieving what can be regarded as near-optimal control performance. 
Consistent with the theoretical predictions of \citeauthor{bergmann}, our strategy effectively suppresses vortex shedding entirely, thereby eradicating the associated drag component \cite{bergmann}. 
Under controlled flow conditions devoid of vortex shedding, the average $C_D$ converges to the range of 1.74 to 1.90, yielding drag reduction rates between 44.2\% and 49.1\%. 
These results highlight the effectiveness of our DRL-based approach in achieving complete vortex suppression.

\subsection{Attaining superior control performance in high Reynolds number regimes}\label{sec:sec4}

As the Reynolds number increases, the complexity of the fluid dynamics also rises, posing a significant challenge in training effective control strategies.
This section undertakes a detailed examination of the control performance achieved via DRL for $Re = 500$, 1000, and 2000. 
The analysis encompasses the efficacy of drag reduction, suppression of lift fluctuations, and the intricacies of the synthetic jet actuation patterns. Furthermore, a comprehensive comparison of the baseline and controlled flow fields—including velocity, pressure, and vorticity distributions—is conducted to elucidate the underlying mechanisms and demonstrate the advanced capabilities of the DRL-based control approach.

\begin{figure*}[htb!]
    \centering
    \includegraphics[width=\textwidth]{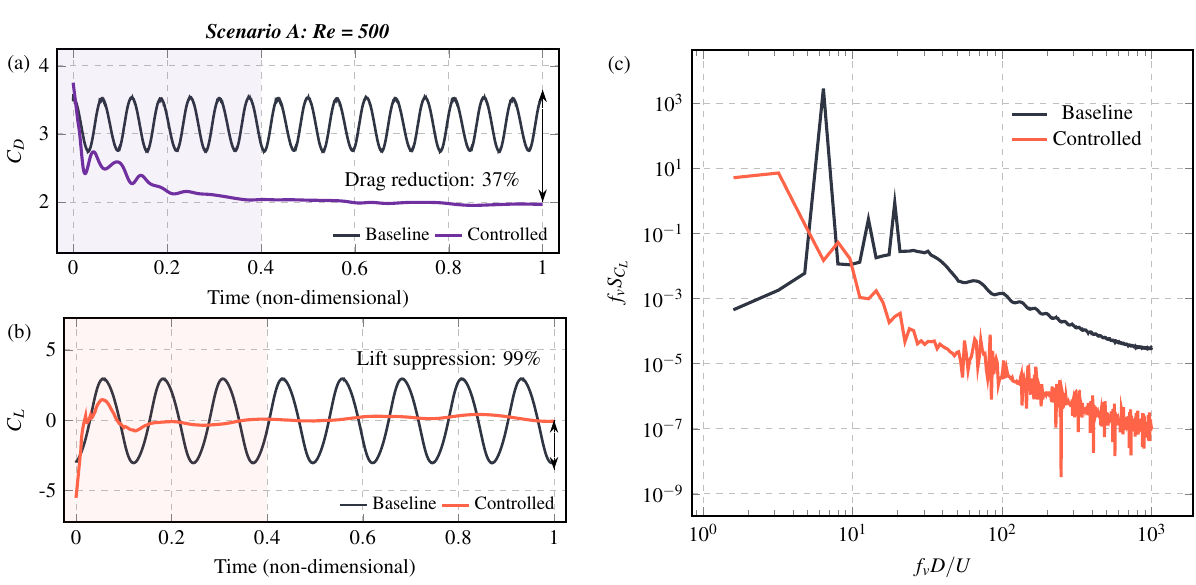}
    \caption{Comparison of control performance for both baseline and controlled flows at $Re = 500$. (a) $C_D$; (b) $C_L$; (c) PSD of $C_L$.}
\label{fig:figure8}
\end{figure*}

\begin{figure*}[htb!]
    \centering
    \includegraphics[width=\textwidth]{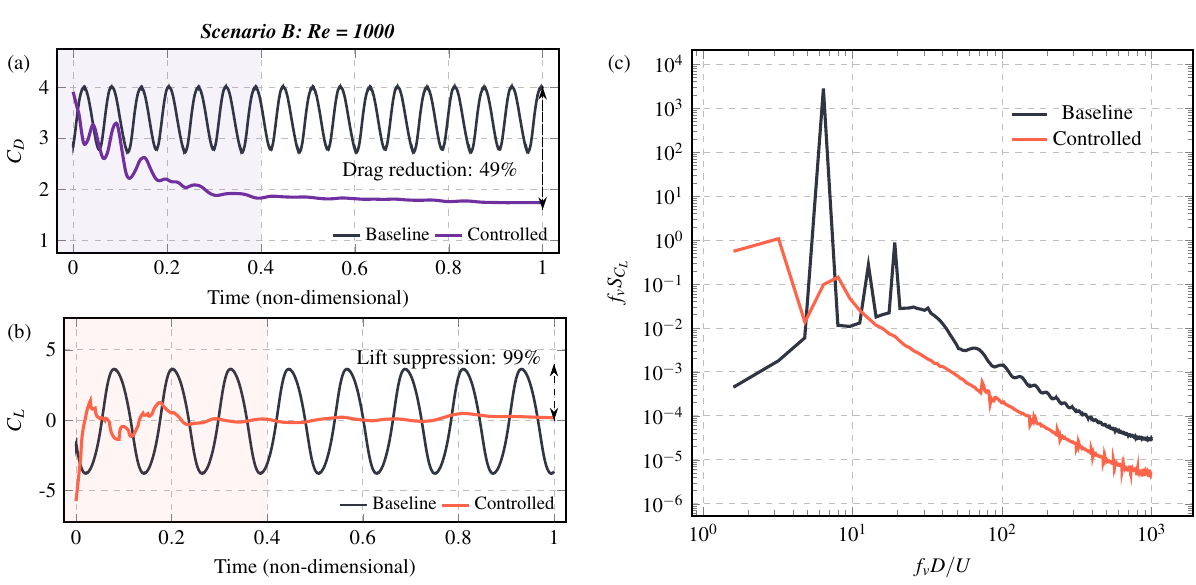}
    \caption{Comparison of control performance for both baseline and controlled flows at $Re = 1,000$. (a) $C_D$; (b) $C_L$; (c) PSD of $C_L$.}
\label{fig:figure9}
\end{figure*}

\begin{figure*}
    \centering
    \includegraphics[width=\textwidth]{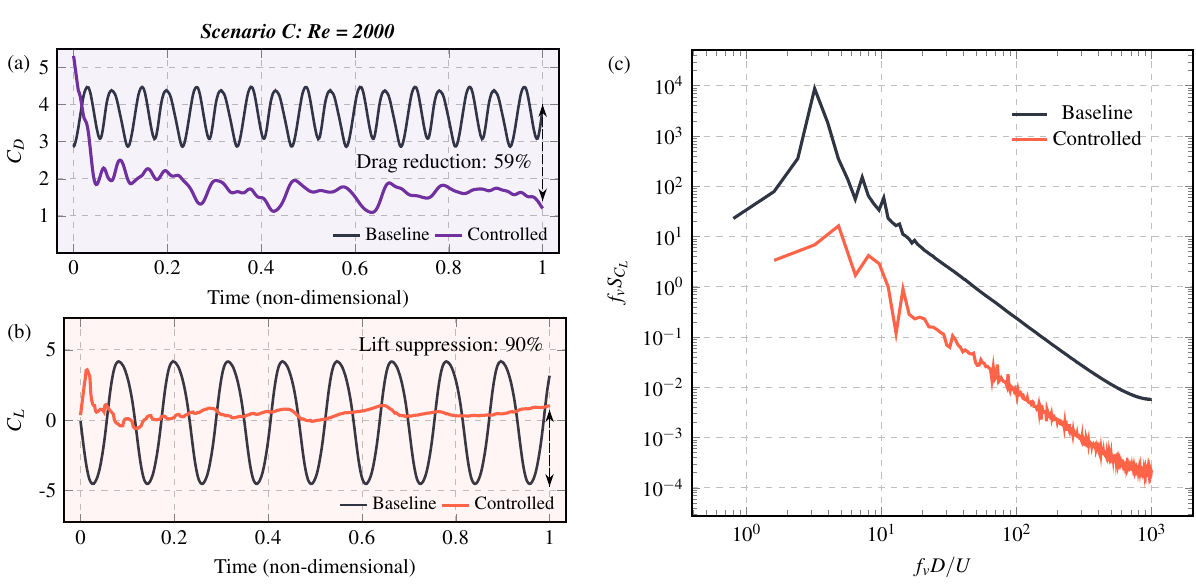}
    \caption{Comparison of control performance for both baseline and controlled flows at $Re = 2,000$. (a) $C_D$; (b) $C_L$; (c) PSD of $C_L$.}
\label{fig:figure10}
\end{figure*}

As illustrated in figure \ref{fig:figure8}, the performance of flow control strategies at $ Re = 500 $ is clearly presented. This includes a comparative analysis of the drag coefficient and lift coefficient for both baseline and controlled flow conditions. Additionally, the results of the Fast Fourier Transform (FFT) applied to the lift coefficient are provided, facilitating a comparison of the dominant frequencies and their associated energy levels in the baseline and controlled flow.
The findings for $ Re = 1000 $ are detailed in figure \ref{fig:figure9}, while the results for $ Re = 2000 $ are depicted in figure \ref{fig:figure10}.

At $ Re = 500 $, figure \ref{fig:figure8} (a) and (b) show that the DRL-based flow control strategy successfully suppresses the periodic oscillations of both the lift and drag coefficients. Compared to the baseline, a 37\% reduction in drag and a 99\% suppression of lift coefficient oscillations are observed under controlled conditions. This demonstrates that the control strategy effectively stabilizes the flow and mitigates undesirable lift fluctuations.
Figure \ref{fig:figure8} (c) presents a power spectral density (PSD) of the lift coefficient for the cylinder with and without active flow control, revealing the dominant frequencies of the lift signal in both the baseline and controlled flows. In the baseline flow, the power spectrum shows clear peaks, indicating significant vortex shedding at those frequencies, which contributes most of the energy to the fluctuations. In contrast, when active flow control is applied, the peaks in the power spectrum of the lift coefficient disappear, indicating that the jet-driven flow has completely disrupted the regular vortex shedding. The controlled flow exhibits a substantial reduction in spectral power over a broad range of frequencies, particularly at the dominant vortex shedding frequency, confirming the effectiveness of the control in suppressing flow instabilities.

The $C_D$, $C_L$, and the PSD of $C_L$ for both the baseline and controlled flows at $Re = 1000$ are presented as shown in figure~\ref{fig:figure9}.
Under controlled conditions, the drag coefficient is reduced by 49\%, and lift coefficient fluctuations are suppressed by 99\%. This control performance is similar to the results observed at $ Re = 500 $. Similarly, the PSD of the lift coefficient for the cylinder also shows comparable control mechanisms. 
Figure \ref{fig:figure10} presents the control performance at $ Re = 2000 $.  The control strategy reaches its highest drag reduction, achieving a remarkable 58.8\%. Despite the presence of a small residual oscillation in $ C_L $, the strategy achieves 90\% lift suppression, indicating that while lift stabilization becomes slightly less effective at higher Reynolds numbers, the overall flow control remains highly efficient.  
Drag reduction improves notably with increasing Reynolds numbers, highlighting the scalability and effectiveness of the control strategy.
Comparing the PSD of the baseline and controlled flows, the controlled flow shows a significant reduction in spectral power over a broad frequency range, with a substantial decrease in peak energy at the dominant frequency. This indicates that, despite significant improvement, vortex shedding at the dominant frequency still persists in the controlled flow at $ Re = 2000 $.

A comparative analysis of the flow control performance between our study and previously reported works at $Re = 500$, $Re = 1,000$, and $Re = 2,000$ is provided in table \ref{tab:table5}.
At $Re = 500$, our configuration at azimuthal positions $90^\circ$ and $270^\circ$ achieves a drag reduction of 37.4 \%, significantly surpassing the 32.2 \% reported by \citeauthor{wang2024dynamic} \cite{wang2024dynamic} for the same jet placement. In addition, our results exhibit much smaller standard deviations in both the drag and lift coefficients, reflecting exceptional flow stability. 
At $Re = 1,000$, the performance gap becomes even more pronounced. Our study achieves a drag reduction of 49.1 \% at $90^\circ$ and $270^\circ$, outperforming the 46.5 \% reported by \citeauthor{wang2024dynamic} \cite{wang2024dynamic}. 
Moreover, the stability parameters of the controlled flow in our results are an order of magnitude lower than those reported in previous studies \cite{wang2024dynamic}. 
Importantly, we achieved complete suppression of vortex shedding at $Re = 500$ and $Re = 1,000$, a level of performance not attained in prior studies, which report only partial suppression under similar conditions \cite{wang2024dynamic,Varela2022,renApplying}.

\begin{table*}[t]
\centering
\caption{Comparative analysis of flow control performance across studies.}
\begin{tabularx}{\textwidth}{
  >{\centering\arraybackslash}p{0.12\textwidth}
  >{\centering\arraybackslash}p{0.15\textwidth}
  >{\centering\arraybackslash}p{0.18\textwidth}
  >{\centering\arraybackslash}p{0.11\textwidth}
  >{\centering\arraybackslash}p{0.11\textwidth}
  >{\centering\arraybackslash}p{0.15\textwidth}
}
\toprule
\toprule
$Re$ & Work & Drag reduction (\%) & $C_\text{D, std}$ & $C_\text{L, std}$ & Suppress Vortex  \\
\midrule
500  & Our work  & 37.4   & 0.015   & 0.148  & Completely  \\
500  & Our work  & 37.2   & 0.019   & 0.208  & Completely   \\
500  & \citet{wang2024dynamic} & 32.2 & 0.252 & 1.610 & Partially \\
1000 & Our work  & 49.1   & 0.034   & 0.062  & Completely  \\
1000 & Our work  & 48.4   & 0.024   & 0.059  & Completely  \\
1000 & \citet{wang2024dynamic}  & 46.5  & 0.310   & 1.610 & Partially  \\
1000 & \citet{Varela2022}  & 30.0   & - & - & Partially \\
1000 & \citet{renApplying} & 30.0   & - & - & Partially  \\
2000 & Our work  & 58.8   & 0.268   & 0.743  & Partially   \\
2000 & \citet{Varela2022}  & 17.7   & -   & -  & Partially  \\
\bottomrule
\bottomrule
\end{tabularx}
\label{tab:table5}
\end{table*}

At a $Re$ of 2000, \citeauthor{Varela2022} achieved only a 17.7\% reduction in drag \cite{Varela2022}. In contrast, our configuration achieved a remarkable 58.8\% reduction in drag, effectively demonstrating the superior performance of our flow control algorithm under high Reynolds number conditions.
A comparison of the controlled flow structures reveals that our method effectively suppresses vortex shedding in the cylinder's wake, significantly outperforming the results reported by \citeauthor{Varela2022}\cite{Varela2022} 
Moreover, the standard deviations of both drag and lift coefficients in our results remain tightly regulated, underscoring the stability and reliability of our synthetic jet control strategy.
Overall, the results clearly underscore the superiority of our study in delivering optimal flow control performance across all Reynolds numbers. 
Our findings demonstrate significant advantages in drag reduction, flow stability, and vortex shedding suppression compared to previous works, affirming the effectiveness and robustness of our optimized synthetic jet configurations.

\begin{figure*}[htb!]
    \centering
    \includegraphics[width=0.99\textwidth]{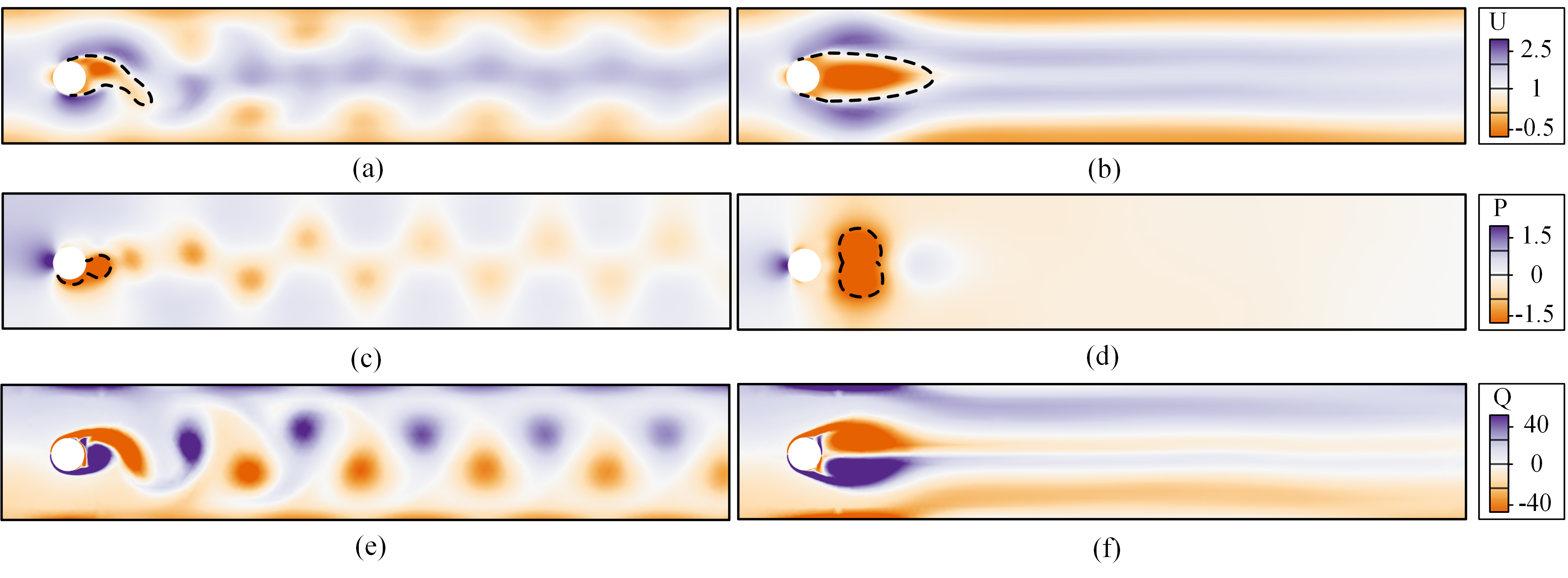}
    \caption{Visualization of instantaneous fields at $Re = 500$. (a) and (b) show the velocity field; (c) and (d) show the pressure field; (e) and (f) show the vorticity field. The left panels depict the baseline flow, while the right panels depict the controlled flow.}
\label{fig:figure11}
\end{figure*}

\begin{figure*}[htb!]
    \centering
    \includegraphics[width=0.99\textwidth]{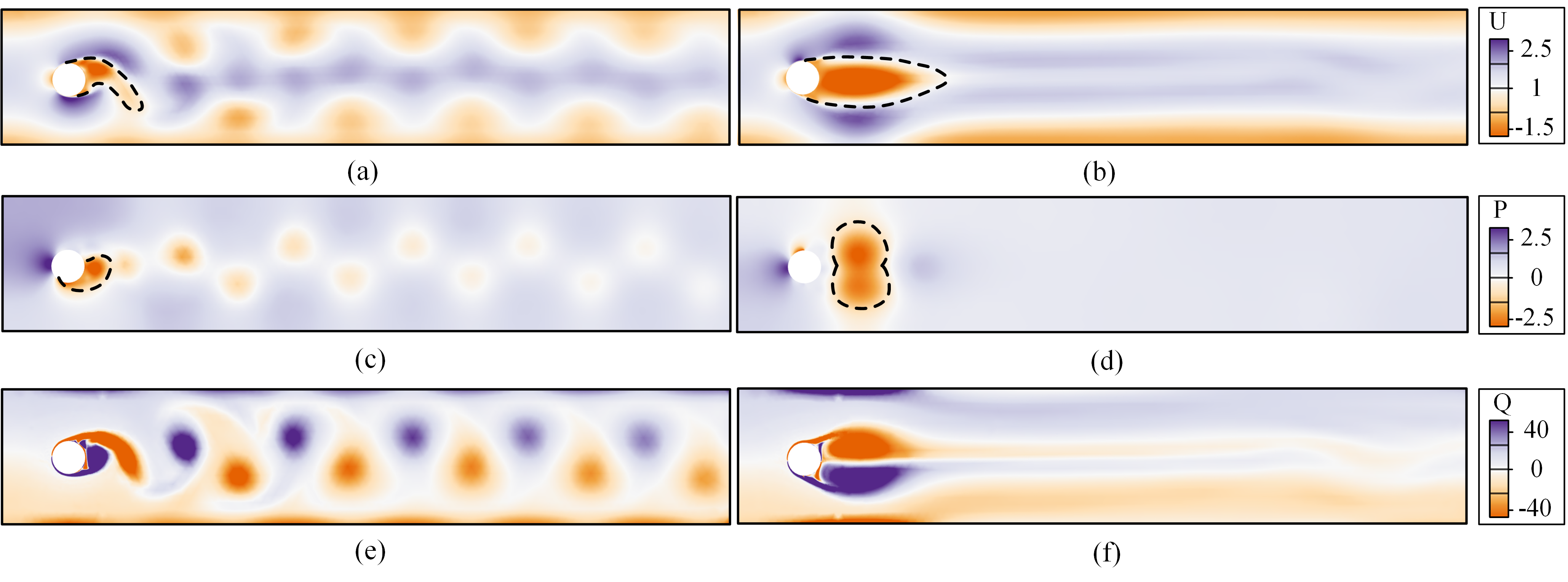}
    \caption{Visualization of instantaneous fields at $Re = 1,000$. (a) and (b) show the velocity field; (c) and (d) show the pressure field; (e) and (f) show the vorticity field. The left panels depict the baseline flow, while the right panels depict the controlled flow.}
    \label{fig:figure12}
\end{figure*}

\begin{figure*}[htb!]
    \centering
    \includegraphics[width=0.99\textwidth]{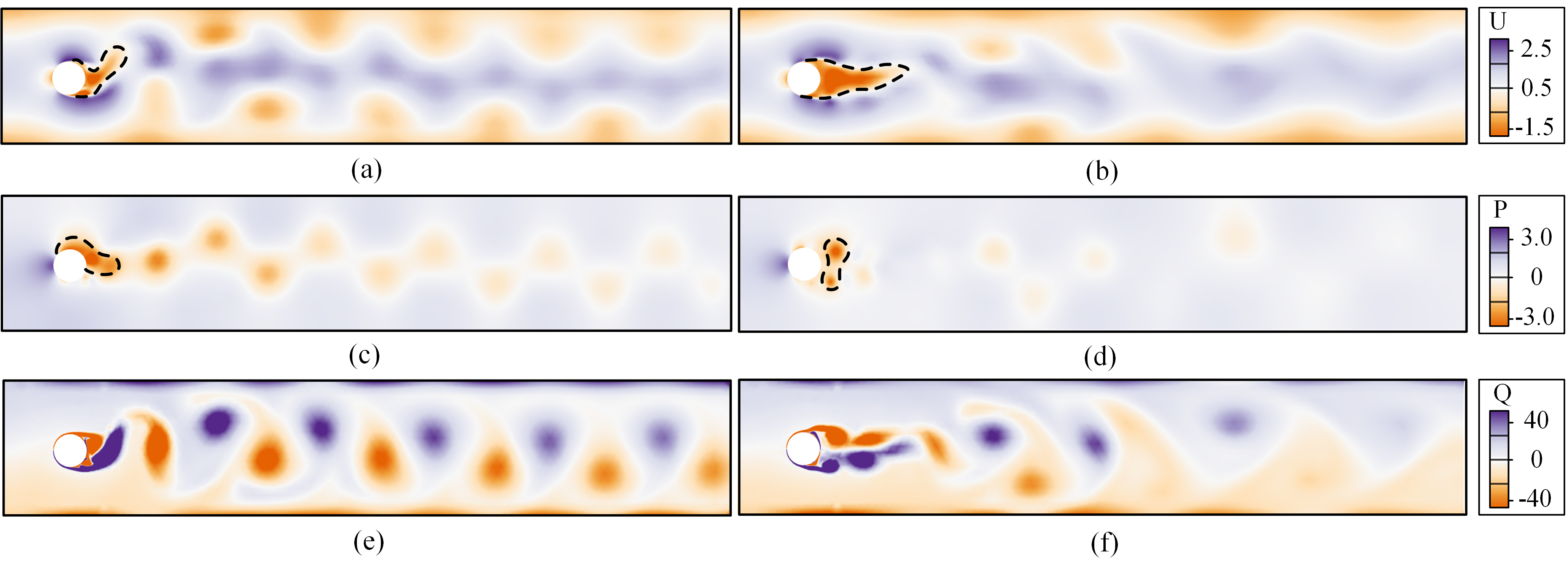}
    \caption{Visualization of instantaneous fields at $Re = 2,000$. (a) and (b) show the velocity field; (c) and (d) show the pressure field; (e) and (f) show the vorticity field. The left panels depict the baseline flow, while the right panels depict the controlled flow.}
    \label{fig:figure13}
\end{figure*}

The flow structures observed before and after the application of control clearly demonstrate the effectiveness of the DRL-based AFC strategy across a range of Reynolds numbers, specifically $ Re = 500 $, $ 1,000 $, and $ 2,000 $. 
Representative flow fields at $Re = 500$, including the streamwise velocity $U$, pressure $P$, and vorticity $Q$, are shown in figure~\ref{fig:figure11} to highlight the contrast between the baseline and controlled cases.
In the absence of control, the velocity field exhibits a well-developed von Kármán vortex street in the wake, characterized by alternating vortex shedding stemming from boundary layer separation on both sides of the cylinder. This results in a broad, unsteady wake region with high levels of pressure fluctuation and momentum deficit. 
The corresponding pressure field further reveals a significant pressure drop across the wake, accompanied by asymmetric distributions indicative of unbalanced lift forces.
Upon activation of the DRL-based control strategy, the flow undergoes a marked topological transition. Vortex shedding is substantially attenuated, leading to a more symmetric and narrowed wake profile. The synthetic jets act to delay boundary layer separation and stabilize the near-wake shear layers, thereby mitigating large-scale vortex formation. The pressure field under control conditions becomes more symmetric, with reduced pressure drag and lower fluctuating lift forces.

The vorticity field further elucidates the impact of flow control. In the absence of actuation, the wake is characterized by strong, periodic vortical structures indicative of a well-developed Kármán vortex street. In contrast, under active control, the vorticity magnitude is significantly reduced, and the wake exhibits streamwise-elongated structures with diminished transverse fluctuations, reflecting a substantial suppression of large-scale unsteadiness.
The effectiveness of synthetic jet actuation arises from its ability to introduce zero-net-mass-flux momentum into the near-wake region in a spatially and temporally controlled manner. Consistent with previous findings \cite{amitay2012modification, Amitay1998Synthetic}, such actuation dynamically alters the local velocity gradients near the separation point, thereby delaying boundary layer separation. As a result, the wake becomes more symmetric and less energetic, contributing to drag reduction and enhanced aerodynamic performance.

A detailed visualization of the flow fields around the cylinder at $ Re = 1,000 $ under baseline and controlled conditions is provided in figure \ref{fig:figure12}. 
The baseline flow is characterized by periodic vortex shedding.
In contrast, the controlled flow exhibits significant suppression of instabilities, featuring a streamlined wake structure, reduced pressure fluctuations, and minimized vorticity intensity.
The recirculation zone, quantified as the region downstream of the cylinder with a negative horizontal velocity component, becomes more stable and organized, reflecting the effectiveness of the control strategy in mitigating unsteady wake dynamics.
These results underscore the efficacy of the control strategy in reducing wake instabilities and stabilizing the flow field around the cylinder at $ Re = 1,000 $, demonstrating the potential for efficient flow control at higher $ Re $.

In the study by \citeauthor{wang2024dynamic}, the DRL-based control strategy led to only partial suppression of vortex shedding at $Re = 500$ and $1{,}000$, which constrained the achievable drag reduction \cite{wang2024dynamic}. In contrast, our approach fully eliminates unsteady wake dynamics, resulting in markedly higher drag reduction and demonstrating superior control effectiveness.
Furthermore, in their study at $Re = 10{,}000$, \citeauthor{wang2024dynamic} achieved a 28.6\% drag reduction in a three-dimensional turbulent wake. This result underscores two critical points \cite{wang2024dynamic}. Firstly, it demonstrates the viability of DRL-based AFC strategies in managing complex 3D turbulent flows. Secondly, it highlights the escalating difficulty of attaining optimal control performance at higher Reynolds numbers, where only partial suppression of flow instabilities was achieved.

As the Reynolds number increases to $ Re = 2000 $, the flow control performance shows a noticeable decline, as illustrated in figure \ref{fig:figure13}. Under controlled conditions, the flow exhibits partial stabilization, but the effectiveness of the control strategy is reduced compared to lower $ Re $. While some suppression of vortex shedding is observed, residual instabilities remain evident, with the velocity field displaying a somewhat streamlined structure but persistent oscillations and reduced elongation of the wake compared to $ Re = 500 $ and $ Re = 1,000 $. The pressure field shows a more uniform distribution under control, with reduced intensity of low-pressure regions in the wake, yet fluctuations remain pronounced, indicating challenges in suppressing pressure instabilities at $ Re = 2000 $. 
The vorticity field shows that the alternating vortices are partially suppressed, but strong residual structures still exist, highlighting the challenges of effective flow control at higher Reynolds numbers.

\citeauthor{Suarez2024} employed DRL to develop active flow control strategies for a three-dimensional circular cylinder at a Reynolds number of 3900 \cite{Suarez2024}. The achieved control performance led only to partial drag reduction and incomplete suppression of vortex shedding, highlighting the inherent challenges of active flow control at moderate to high Reynolds numbers.
Their findings exhibit strong consistency with the present study, as the DRL agent reduces drag by enlarging the recirculation region and suppressing mixing between the inner and outer shear layers, which in turn stabilizes and extends the outer shear layer and reshapes the wake into a more aerodynamically favorable configuration. In addition, their work confirms the feasibility of DRL-based control strategies in three-dimensional flow regimes.

\section{CONCLUSIONS}\label{sec:Conclusions} 

We study investigated the performance of DRL-based AFC strategies at intermediate and high Reynolds numbers. The DRL-trained control policies successfully achieved multiple objectives, including minimizing the mean drag coefficient, suppressing fluctuations in both drag and lift coefficients, and mitigating vortex shedding in the wake of the cylinder. Additionally, the placement of synthetic jets was found to significantly influence control effectiveness and energy efficiency, emphasizing the importance of optimal configurations. The key findings are summarized as follows:

\begin{enumerate}
    \item \textit{Impact of jet placement on control performance and energy efficiency.}
    Across various Reynolds numbers, the flow control performance when synthetic jets were placed at configurations $ J_{56} $ and $ J_{78} $ consistently outperformed other setups. This was reflected in achieving the highest drag reduction rates, optimal lift control, and the most evident stability in the controlled flow. In contrast, the configuration at $ J_{910} $ exhibited the weakest performance, with limited drag reduction and significant fluctuations in the controlled flow. The control strategies generated under different jet configurations indicated that synthetic jets positioned at locations $ J_{56} $ and $ J_{78} $ produced smoother and more stable control signals. This suggests that among the two jet configurations, flow control was more energy-efficient and stable. In comparison, control strategies with synthetic jets located at $ J_{12} $ and $ J_{910} $ showed the highest instability initially and required the highest and most unstable external energy consumption to maintain control performance in the mid-to-late stages of control.
    These findings highlight the critical role of injection position in optimizing flow control performance and energy efficiency, while also demonstrating the real-time and intelligent nature of the control strategy obtained through DRL-based training.
    
    \item \textit{Achieving multi-objective flow control across multiple $Re$.}
    The DRL-based AFC strategy demonstrated robust performance at intermediate Reynolds numbers but faced challenges at higher $Re$. At $Re = 500$, the strategy achieved a 37.4\% drag reduction and suppressed lift oscillation amplitudes by 99\%, resulting in a fully stabilized wake with reduced recirculation zones, uniform pressure distribution, and weakened vorticity structures. At $Re = 1,000$, the drag reduction increased to 49.1\%, with near-complete elimination of lift oscillations and further enhanced wake stability. However, at $Re = 2,000$, the strategy achieved a 58.8\% drag reduction and a 90\% suppression of lift oscillations, but some residual wake instabilities persisted. These results indicate the strategy’s strong performance at intermediate Reynolds numbers while highlighting its limitations under turbulent conditions, where increasing flow complexity poses significant challenges.
    
    \item \textit{Analysis of the physical mechanism of synthetic jets in flow control.}
    The activation of synthetic jets on a blunt body significantly alters the flow characteristics, forming closed recirculation zones near the jets or around the cylinder. The formation of these recirculation zones enables the synthetic jets to exert substantial overall effects on scales much larger than the characteristic length of the jets themselves, thereby influencing the cylinder’s overall lift and drag. Furthermore, a sensitivity analysis of synthetic jet placement demonstrates that positioning the jets near the flow separation region is particularly advantageous for achieving optimal control performance.
\end{enumerate}

The results of this study highlight the efficacy of DRL-based AFC strategies in achieving significant flow control improvements, particularly at intermediate Reynolds numbers. The findings emphasize the importance of optimal jet placement and the scalability of these strategies for higher Reynolds numbers, despite the challenges posed by increased flow complexity. 
Future work will extend this framework to three-dimensional flows, where spanwise instabilities introduce additional control challenges, and to higher Reynolds number regimes, where increased turbulence demands more robust and scalable control policies. These extensions are essential for advancing DRL-based AFC toward practical aerodynamic applications.

\section*{ACKNOWLEDGMENTS}

This research is supported by the Ministry of Industry and Information Technology, China, under the project "Research on Digital Reconstruction of Ship Final Assembly Construction" (Grant No. CBG01N23-01-02).

\section*{AUTHOR DECLARATIONS}


The authors report no conflict of interest.



\section*{DATA AVAILABILITY}

The data that support the findings of this study are available from the corresponding author upon reasonable request.



\bibliographystyle{elsarticle-num-names} 
\bibliography{elsarticle}

\end{document}